\documentclass[runningheads]{llncs}

\usepackage{amsmath,amssymb} 

\usepackage{amsthm}

\usepackage{citesort}
\usepackage{url}
\usepackage{color}
\usepackage{xcolor}
\usepackage{listings}
\usepackage{booktabs}
\usepackage{multicol}
\usepackage{multirow}
\usepackage{graphicx}
\usepackage{makecell}

\definecolor{mygray}{rgb}{0.98,0.98,0.98}

\usepackage{tikz-cd}
\usepackage{multicol}
\usepackage{wrapfig}

\newcommand{\NI}{\noindent}

\newcommand{\AND}{\wedge}
\newcommand{\BIGAND}{\bigwedge}
\newcommand{\BIGOR}{\bigvee}
\newcommand{\OR}{\vee}
\newcommand{\X}{\mathsf{X}}
\newcommand{\F}{\mathsf{F}}
\newcommand{\G}{\mathsf{G}}
\newcommand{\U}{\mathbin{\mathsf{U}}}

\newcommand{\RESP}{resp.,\ } 
\newcommand{\WRT}{w.r.t.\ } 
\newcommand{\IE}{\textit{i.e.},\ } 
\newcommand{\EG}{e.g.,\ } 
\newcommand{\STARTEG}{E.g.,\ } 
\newcommand{\CF}{cf.\ } 

\newenvironment{TABLE}{\begin{table}[t]\begin{center}}{\end{center}\vspace{-10mm}\end{table}}

\newcommand{\NAT}{\mathbb{N}}
\newcommand{\BOOL}{\mathbb{B}}
\newcommand{\SUBFORMULAE}[1]{\mathsf{sf}(#1)}

\newcommand{\SCARLET}[1]{${\sc Scarlet}$(#1)}
\newcommand{\SIMPLE}[1]{${\sc BenchBase}$(#1)}
\newcommand{\SAMPLEBENCH}[1]{${\sc Sampling}$(#1)}
\newcommand{\HAMMINGBENCH}[1]{${\sc Hamming}$(#1)}
\newcommand{\HAMMINGDISTANCE}[2]{\mathsf{hamm}(#1, #2)}
\newcommand{\SUFFIX}[1]{\mathsf{sc}(#1)}
\newcommand{\SUFFIXNE}[1]{\mathsf{sc}^{+}(#1)}

\newenvironment{FIGURE}{\begin{figure}[t] \centering}{\vspace{-4mm}\end{figure}}
\newcommand{\BASENOARGS}{\textsc{Enum}}
\newcommand{\BASE}[3]{\BASENOARGS(#1, #2, #3)}

\newcommand{\OVERFIT}[1]{\mathsf{overfit}(#1)}
\newcommand{\EMPH}[1]{\emph{#1}}
\newcommand{\STRING}[1]{\langle #1 \rangle}
\newcommand{\LANG}[1]{\mathsf{lang}(#1)}
\newcommand{\SIZE}[1]{|\!|#1|\!|}
\newcommand{\CARD}[1]{\# #1}
\newcommand{\LENGTH}[1]{|\!|#1|\!|}
\newcommand{\COST}[1]{\mathsf{cost}(#1)}
\newenvironment{GRAMMAR}{\[\begin{array}{lcl}}{\end{array}\]}

\newcommand{\VERTICAL}{\  \mid\hspace{-3.0pt}\mid \ }
\newcommand{\TRACES}[1]{\mathsf{traces}(#1)}
\newcommand{\BENCHURL}[1]{\url{https://github.com/martinberger/ltl-learning/blob/master/bench/#1}}

\newcommand{\PROGRAM}[1]{\mathsf{#1}}

\newcommand{\TRUE}{\PROGRAM{true}}
\newcommand{\FALSE}{\PROGRAM{false}}

\usepackage{ifthen}
\newboolean{showcomments}
\setboolean{showcomments}{true}
\ifthenelse{\boolean{showcomments}}
  {\newcommand{\mynote}[2]{
    \fbox{\bfseries\sffamily\scriptsize#1}
    {\small$\blacktriangleright$\textsf{\emph{#2}}$\blacktriangleleft$}
   }
  }
  {\newcommand{\mynote}[2]{}
  }

\newcommand{\CHARACTERISTIC}[2]{\mathbf{1}^{#1}_{#2}}
\newcommand{\POWERSET}[1]{\mathfrak{P}(#1)}
\newcommand{\LTLFORMULA}[1]{\mathsf{LTL}(#1)}

\newcommand{\HAMMING}[2]{\mathsf{Hamm}(#1, #2)}

\newcommand{\SYNTH}[1]{\mathsf{learn}(#1)}
\newcommand{\CODERAW}[1]{
  \begingroup
  \lstset{basicstyle=\ttfamily\small,language=Python} 
  \lstinline!#1!
  \endgroup
}
\newcommand{\CODE}[1]{
\!\!\!\!\!\!\CODERAW{#1}\!\!\!\!\!
  }

\newcommand{\STARTRANDOMSPLIT}{\mathsf{randSplit}}
\newcommand{\RANDOMSPLIT}[1]{\STARTRANDOMSPLIT(#1)}
\newcommand{\DETERMNINISTICSPLIT}[1]{\mathsf{detSplit}(#1)}

\newcommand{\RANDOMAUX}[1]{\mathsf{aux}(#1)}

\newcommand{\CONG}{\simeq}
\newcommand{\NOVSPACEPARAGRAPH}[1]{\NI\textbf{\emph{#1}.}}
\newcommand{\PARAGRAPH}[1]{\vspace{2mm}\NOVSPACEPARAGRAPH{#1}}
\newcommand{\FS}{\rightarrow}

\begin{document}



 \title{LTL learning on GPUs}

\author{Mojtaba Valizadeh\inst{1, 2} \and Nathana\"el Fijalkow\inst{3} \and Martin Berger\inst{1, 4}}
\authorrunning{M. Valizadeh et al.}
\institute{
  University of Sussex, Brighton, UK, \\
  \and Neubla UK Ltd, \email{Valizadeh.Mojtaba@gmail.com} \\
  \and CNRS, LaBRI and Universit\'e de Bordeaux, France, \email{nathanael.fijalkow@gmail.com} \\
  \and Montanarius Ltd, London, UK, \email{contact@martinfriedrichberger.net}
  }

\maketitle              

\begin{abstract}
  Linear temporal logic (LTL) is widely used in industrial
  verification.  LTL formulae can be learned from 
  traces. Scaling LTL formula learning is an open problem. We
  implement the first GPU-based LTL learner using a novel form of
  enumerative program synthesis. The learner is sound and
  complete. Our benchmarks indicate that it handles traces at least
  2048 times more numerous, and on average at least 46 times faster
  than existing state-of-the-art learners. This is achieved with, among others, 
  novel branch-free LTL semantics that has $O(\log n)$ time
  complexity, where $n$ is trace length, while previous
  implementations are $O(n^2)$ or worse (assuming bitwise boolean
  operations and shifts by powers of 2 have unit costs---a realistic
  assumption on modern processors).


\end{abstract}

\section{Introduction}

Program verification means demonstrating that an implementation
exhibits the behaviour required by a specification. But where do
specifications come from? Handcrafting specifications does not scale.
One solution is automatically to \EMPH{learn} them from example runs
of a system.  This is sometimes referred to as trace analysis. A
trace, in this context, is a sequence of events or states captured
during the execution of a system.  Once captured, traces are often
converted into a form more suitable for further processing, such as
finite state automata or logical formulae.  Converting traces into
logical formulae can be done with program synthesis.  Program
synthesis is an umbrella term for the algorithmic generation of
programs (and similar formal objects, like logical formulae) from
specifications, see \cite{DavidC:prosynsao,GulwaniS:prosyn} for an
overview.  Arguably, the most popular logic for representing traces is
linear temporal logic (LTL) \cite{PnueliA:temlogop}, a modal logic for
specifying properties of finite or infinite traces.  The
\EMPH{LTL learning problem} idealises the algorithmic essence of
learning specifications from example traces, and is given as follows.

\begin{itemize}

\item \textbf{Input:} Two sets $P$ and $N$ of traces over a fixed
  alphabet. 
   
\item \textbf{Output:} An LTL formula $\phi$ that is (i) \EMPH{sound}:
  all traces in $P$ are accepted by $\phi$, all traces in $N$ are
  rejected by $\phi$; (ii) \EMPH{minimal}, meaning no strictly smaller
  sound formula exists.

\end{itemize}
When we weaken minimality to minimality-up-to-$\epsilon$, we speak of
\EMPH{approximate} LTL learning. Both forms of LTL learning are
NP-hard \cite{FijalkowN:compolealtlffe,MascleC:learemffeih}.  A
different and simpler problem is \EMPH{noisy} LTL learning, which is
permitted to learn unsound formulae, albeit only up-to a give
error-rate.

LTL learning is an active research area in software engineering,
formal methods, and artificial
intelligence~\cite{NeiderG18,RahaR:scaanyaflfoltl,LemieuxC:invprobuttlsm,LemieuxC:genltlsm,AmmonsG:minspe,GabelM:onlinfaeotp,GabelM:symminots,GabelM:javfulamoftpfdt,WeimerW:mintemsfed,YangJ:permtarfit,YoganandaJeppuN:lersymafset,JeppuNY:leaconmflet,CamachoA:leaintmeiltl,Camacho_McIlraith_2019,ijcai2019-0776,IeloA:towilpblpl,GaglioneNRTX22,PengB:purminlsfitit,LuoL:briltlitgifllf}.
We refer to~\cite{Camacho_McIlraith_2019} for a longer discussion.
Many approaches to LTL learning have been explored.
One common and natural method involves using search-based program
synthesis, often paired with templates or sketches, such as parts of
formulas, automata, or regular expressions. Another
leverages SAT solvers.  LTL learning is also being pursued using
Bayesian inference, or inductive logic programming.  Learning
specifically tailored small fragments of LTL often yields the best results
in practice~\cite{RahaR:scaanyaflfoltl}.  Learning from noisy data is
investigated in~\cite{GaglioneNRTX22,PengB:purminlsfitit,LuoL:briltlitgifllf}.
All have in common is that they don't scale, and have not been optimised for GPUs.
Traces arising in industrial practice are commonly long (millions of
characters), and numerous (millions of traces).  Extracting useful
information automatically at such scale is currently a major problem,
\EG the state-of-the-art learner in \cite{RahaR:scaanyaflfoltl} cannot reliably learn
formulae greater than size 10.  This is less than ideal.  Our aim is
to change this.

Graphics Processing Units (GPUs) are the work-horses of
high-performance computing. The acceleration they provide to
applications compatible with their programming paradigm can surpass
CPU performance by several orders of magnitude, as notably evidenced
by the advancements in deep learning. A significant spectrum of
applications, especially within automated reasoning—like SAT/SMT
solvers and model checkers—has yet to reap the benefits of GPU
acceleration.  In order for an application to be “GPU-friendly”, it
needs to have high parallelism, minimal data-dependent branching, and
predictable data movement with substantial data locality
\cite{DallyWJ:domspeha,Hennessy17Computer,HwuWMW:promaspp}. Current automated
reasoning algorithms are predominantly branching-intensive and appear
sequential in nature, but it is unclear whether they are inherently
sequential, or can be adapted to GPUs.
\begin{quote}
  \textbf{Research question.}  \EMPH{Can we scale LTL learning to at least
  1000 times more traces without sacrificing trace length, learning
  speed or approximation ratio (cost increase of learned formula over
  minimum) compared to existing work, by employing suitably adapted
  algorithms on a GPU?}
\end{quote}
We answer the RQ in the affirmative by developing the first
GPU-accelerated LTL learner.  Our work takes inspiration from
\cite{ValizadehM:seabasreioag}, the first GPU-accelerated minimal
regular expression inferencer.  Scaling has two core orthogonal
dimensions: more traces, and longer traces.  We solve one problem
\cite{ValizadehM:seabasreioag} left open: scaling to
more traces. Our key decision, giving up on learning minimal formula
while remaining sound and complete, enables two principled
algorithmic techniques.
\begin{itemize}

\item \textbf{Divide-and-conquer (D\&C).}  If a learning task has
  too many traces, split it into smaller specifications, learn those
  recursively, and combine the learned formulae using logical
  connectives.

\item \textbf{Relaxed uniqueness checks (RUCs).} Often
  generate-and-test program synthesis caches synthesis results to
  avoid recomputation.  \cite{ValizadehM:seabasreioag} granted cache
  admission only after a uniqueness check.  We relax uniqueness
  checking by  (pseudo-)randomly rejecting some unique formulae.

\end{itemize}
In addition, we design novel algorithms and data structures,
representing LTL formulae as contiguous matrices of bits.  This allows
a GPU-friendly implementation of all logical operations with linear
memory access and suitable machine instructions, free from
data-dependent branching.  Both D\&C and RUCs may lose minimality and
are thus unavailable to \cite{ValizadehM:seabasreioag}, see Appendix
\ref{appendix_comparison} for a comparison.  Our benchmarks
show that the approximation ratio is typically small.

\PARAGRAPH{Contributions}
In summary, our contributions are as follows:

\begin{itemize}

\item A new enumeration algorithm for LTL learning, with a
  branch-free implementation of LTL semantics that is
  $O(\log n)$ in trace length (assuming unit cost for logical and
  shift operations).

\item A CUDA implementation of the algorithm, for benchmarking and
  inspection.

\item A parameterised benchmark suite useful for evaluating the
  performance of LTL learners, and a novel methodology for quantifying
  the loss of minimality induced by approximate LTL learning.

\item Performance benchmarks showing that our implementation is both faster,
  and can handle orders of magnitude more traces, than existing work.

\end{itemize}

\section{Formal preliminaries}

We write $\CARD{S}$ for the cardinality of set $S$.  $\NAT = \{0, 1,
2, ...\}$, $[n]$ is for $\{0, 1, ..., n-1\}$ and $[m, n]$ for $\{m, m+1, ..., n-1\}$.
$\BOOL$ is $\{0, 1\}$
where $0$ is falsity and $1$ truth.  $\POWERSET{A}$ is the
\EMPH{powerset} of $A$. The \EMPH{characteristic function} of a set $S$
is the function $\CHARACTERISTIC{A}{S} : A \FS \BOOL$
which maps $a \in A$ to 1 iff $a \in S$. We usually write
$\CHARACTERISTIC{}{S}$ for $\CHARACTERISTIC{A}{S}$.  An
\EMPH{alphabet} is a finite, non-empty set $\Sigma$, the elements of
which are \EMPH{characters}. A \EMPH{string of length $n \in \NAT$}
over $\Sigma$ is a map $w : [n] \FS \Sigma$.  We write $\LENGTH{w}$
for $n$. We often write $w_i$ instead of $w(i)$, and $v \cdot w$, or
just $vw$, for the concatenation of $v$ and $w$, $\epsilon$ for the
empty string and $\Sigma^*$ for all strings over $\Sigma$.  A
\EMPH{trace} is a string over powerset alphabets, \IE
$(\POWERSET{\Sigma})^*$. We call $\Sigma$ the \EMPH{alphabet} of the
trace and write $\TRACES{\Sigma}$ for all traces over $\Sigma$.  A
\EMPH{word} is a trace where each character has cardinality 1. We
abbreviate words to the corresponding strings, \EG $\STRING{\{t\},
  \{i\}, \{n\}}$ to $tin$.  We say $v$ is a \EMPH{suffix} of $w$ if $w
= u v$, and if $\LENGTH{u} = 1$ then $v$ is an \EMPH{immediate}
suffix.  We write $\SUFFIX{S}$ for the \EMPH{suffix-closure} of $S$.
$S$ is \EMPH{suffix-closed} if $\SUFFIX{S} \subseteq S$.
$\SUFFIXNE{S}$ is the \EMPH{non-empty suffix closure} of $S$, \IE
$\SUFFIX{S} \setminus \{\epsilon\}$.  \EMPH{From now on we will speak
  of the suffix-closure to mean the non-empty suffix closure.}  The
\EMPH{Hamming-distance} between two strings $s$ and $t$ of equal
length, written $\HAMMINGDISTANCE{s}{t}$, is the number of indices $i$
where $s(i) \neq t(i)$. We write $ \HAMMING{s}{\delta}$ for the set
$\{t \in \Sigma^*\ |\ \HAMMINGDISTANCE{s}{t} = \delta, \LENGTH{s} =
\LENGTH{t}\}$.

\EMPH{LTL formulae} over $\Sigma = \{p_1, ..., p_n\}$ are given by the
following grammar.
\begin{GRAMMAR}
  \phi
  &\quad ::= \quad\ &
  p \VERTICAL
  \neg \phi \VERTICAL
  \phi \AND \phi \VERTICAL
  \phi \OR \phi \VERTICAL
  \X \phi \VERTICAL
  \F \phi \VERTICAL
  \G \phi \VERTICAL
  \phi \U \phi
\end{GRAMMAR}
The \EMPH{subformulae} of $\phi$ are denoted $\SUBFORMULAE{\phi}$. We
say $\psi \in \SUBFORMULAE{\phi}$ is \EMPH{proper} if $\phi \neq
\psi$. A formula is in \EMPH{negation normal form} (NNF) if all subformulae
containing negation are of the form $\neg p$. It is \EMPH{$\U$-free}
if no subformula is of the form $\phi \U \psi$. We write
$\LTLFORMULA{\Sigma}$ for the set of all LTL formulae over $\Sigma$.
We use $\TRUE$ as an abbreviation for $p \OR \neg p$ and $\FALSE$ for $p
\AND \neg p$.  We call $\X, \F, \G, \U$ the \EMPH{temporal}
connectives, $\AND, \OR, \neg$ the \EMPH{propositional} connectives,
$p$ the \EMPH{atomic} propositions and, collectively name them the
\EMPH{LTL connectives}.  Since we learn from finite traces, we
interpret LTL over finite traces \cite{DeGiacomoG:lintemlaldloft}.
The satisfaction relation $tr, i \models \phi$, where $tr$ is a trace
over $\Sigma$ and $\phi$ from $\LTLFORMULA{\Sigma}$ is standard, here
are some example clauses: $tr, i
\models \X \phi $, if $tr, i+1 \models \phi$, $tr, i \models \F \phi
$, if there is $i \leq j < \LENGTH{tr}$ with $tr, j \models \phi$, and
$tr, i \models \phi \U \phi'$, if there is $i \leq j < \LENGTH{tr}$
such that: $tr, k \models \phi$ for all $i \leq k < j$, and $tr, j
\models \phi'$.  If $i \geq \LENGTH{tr}$ then $tr, i \models \phi$ is
always false, and  $tr \models \phi$ is short for $tr, 0
\models \phi$.

A \EMPH{cost-homomorphism} is a map $ \COST{\cdot}$ from LTL
connectives to positive integers. We extend it to LTL formulae
homomorphically: $\COST{\phi\ op\ \psi} = \COST{op} + \COST{\phi} +
\COST{\psi}$, and likewise for other arities. If $\COST{op} = 1$ for
all LTL connectives we speak of \EMPH{uniform cost}.  So the uniform
cost of $\TRUE$ and $\FALSE$ is 4.  \EMPH{From now on all
  cost-homomorphisms will be uniform, except where stated otherwise.}

A \EMPH{specification} is a pair $(P, N)$ of finite sets of traces
such that $P \cap N = \emptyset$.  We call $P$ the \EMPH{positive}
examples and $N$ the \EMPH{negative} examples.  We say $\phi$
\EMPH{satisfies, separates} or \EMPH{solves} $(P, N)$, denoted $\phi
\models (P, N)$, if for all $tr \in P$ we have $tr \models \phi$, and
for all $tr \in N$ we have $tr \not \models \phi$.  A
\EMPH{sub-specification} of $(P, N)$ is any specification $(P', N')$
such that $P' \subseteq P$ and $N' \subseteq N$. Symmetrically, $(P,
N)$ is an \EMPH{extension} of $(P', N')$.  We can now make the
\EMPH{LTL learning problem} precise:
\begin{itemize}

\item \textbf{Input:} A specification $(P, N)$, and a cost-homomorphism $\COST{\cdot}$.

\item \textbf{Output:} An LTL formula $\phi$ that is \EMPH{sound}, \IE
  $\phi \models (P, N)$, and \EMPH{minimal}, \IE $\psi \models (P, N)$
  implies $\COST{\phi} \leq \COST{\psi}$.

\end{itemize}

\NI Cost-homomorphisms let us influence LTL learning: \EG by assigning
a high cost to a connective, we prevent it from being used in learned formulae.
The \EMPH{language at $i$} of $\phi$, written $\LANG{i, \phi}$, is
$\{tr \in \TRACES{\Sigma}\ |\ tr, i \models \phi\}$.  We write
$\LANG{\phi}$ as a shorthand for $\LANG{0, \phi}$ and speak of the
\EMPH{language} of $\phi$. We say $\phi$ \EMPH{denotes} a language $S
\subseteq \Sigma^*$, \RESP a trace $tr \in \Sigma^*$, if $\LANG{\phi}
= S$, \RESP $\LANG{\phi} = \{tr\}$.  We say two formulae $\phi_1$ and
$\phi_2$ are \EMPH{observationally equivalent}, written $\phi_1 \CONG
\phi_2$, if they denote the same language. Let $S$ be a set of
traces. Then we write
\[
\phi_1 \CONG \phi_2 \mod S \qquad\text{iff}\qquad \LANG{\phi_1} \cap
S = \LANG{\phi_2} \cap S
\]
and say $\phi_1$ and $\phi_2$ are \EMPH{observationally equivalent
  modulo} $S$.  
The following related definitions will be useful later.  Let $(P, N)$
be a specification.  The \EMPH{cardinality of $(P, N)$}, denoted
$\CARD{(P, N)}$, is $\CARD{P} + \CARD{N}$.  The \EMPH{size} of a set
$S$ of traces, denoted $\SIZE{S}$ is $\Sigma_{tr \in S} \LENGTH{tr}$.
We extend this to specifications: $\SIZE{(P, N)}$ is $\SIZE{P} +
\SIZE{Q}$.  The \EMPH{cost} of a specification $(P, N)$, written
$\COST{P, N}$ is the uniform cost of a minimal sound formula for
$(P, N)$.  An extension of $(P, N)$ is \EMPH{conservative} if any
minimal sound formula for $(P, N)$ is also minimal and sound
for the extension.  We note a useful fact: if $\phi$ is a minimal
solution for $(P, N)$, and also $\phi \models (P', N')$ then $(P \cup
P', N \cup N')$ is a conservative extension.

\PARAGRAPH{Overfitting} It is possible to express a trace $tr$, respectively a set $S$ of
traces, by a formula $\phi$, in the sense that $\LANG{\phi} = \{tr\}$,
\RESP $\LANG{\phi} = S$.
We define the function $\OVERFIT{\cdot}$ on sets of characters, traces, sets and specifications as follows.
  \begin{itemize}

  \item $\OVERFIT{\{a_1, ..., a_k\}} = (\BIGAND_{i} a_i) \AND  \BIGAND_{b \in \Sigma \setminus \{a_1, ..., a_k\}} \neg b $.

  \item $\OVERFIT{\epsilon} = \neg \X(\TRUE)$ and  $\OVERFIT{a \cdot tr} = \OVERFIT{a}  \AND \X(\OVERFIT{tr})$.

  \item $\OVERFIT{S} = \BIGOR_{tr \in S} \OVERFIT{tr}$

  \item $\OVERFIT{P, N} = \OVERFIT{P}$

  \end{itemize}
The following are immediate from the definitions: (i) For all
specifications $(P, N)$: $\LANG{\OVERFIT{P, N}} = P$, (ii)
$\OVERFIT{P, N} \models (P, N)$, and (iii) the \EMPH{cost of 
overfitting}, \IE $\COST{\OVERFIT{P, N}}$, is $O(\SIZE{P} +
\CARD{\Sigma})$.  Note that $\OVERFIT{P, N}$ is overfitting only on
$P$, and (ii) justifies this choice.

\section{High-level structure of the algorithm}\label{section_algorithm}

\begin{FIGURE}
\includegraphics[width=.85\linewidth]{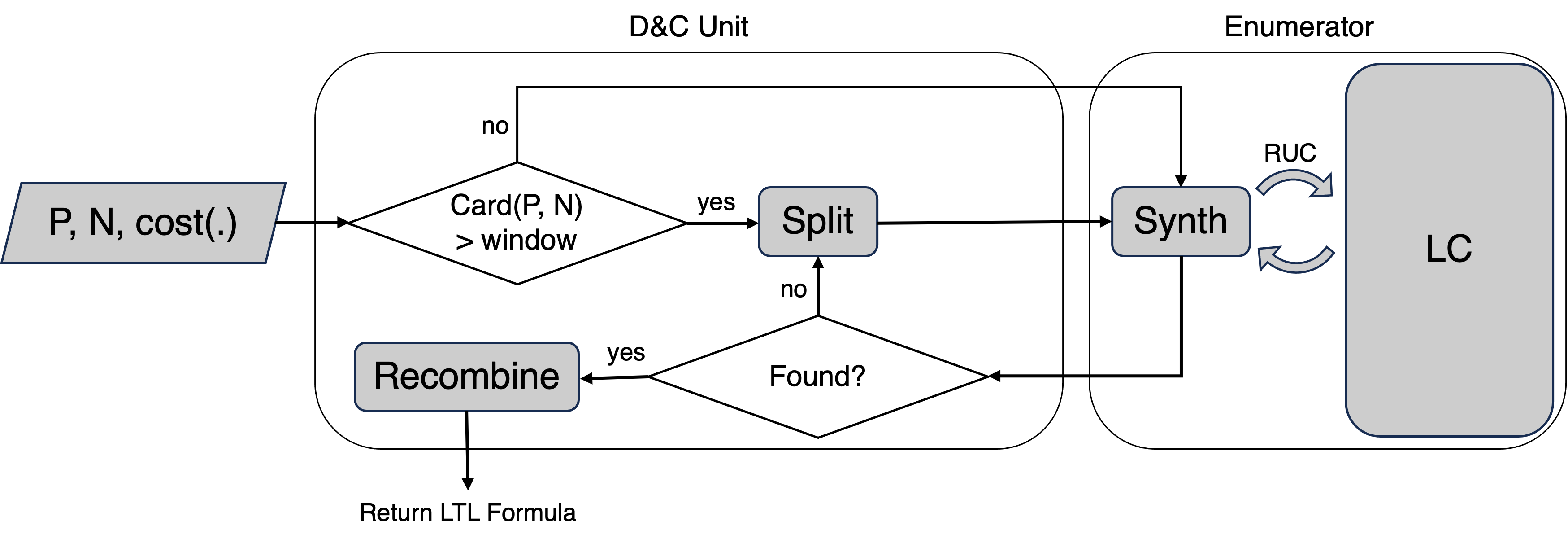}
\caption{High-level structure of our algorithm. LC is short for
  language cache.}\label{image_overall_structure}
\end{FIGURE}

Figure \ref{image_overall_structure} shows the two main parts of our
algorithm: the \EMPH{divide-and-conquer unit}, short D\&C-unit, and
the \EMPH{enumerator}. Currently, only the enumerator is
implemented for execution of a GPU. For convenience, our D\&C-unit
is in Python and runs on a CPU. Implementing the $D \& C$-unit on a
GPU poses no technical challenges and would make our implementation perform
better.

Given $(P, N)$, the D\&C-unit checks if the specification is small
enough to be solved by the enumerator directly.  If not, the
specification is recursively decomposed into smaller
sub-specifications. When the recursive invocations return formulae,
the D\&C-unit combines them into a formula separating $(P, N)$, see
\S \ref{d_and_c} for details.  For small enough $(P, N)$, the
enumerator performs a bottom-up enumeration of LTL formulae by
increasing cost, until it finds one that separates $(P, N)$.  Like the
enumerator in \cite{ValizadehM:seabasreioag}, our enumerator uses a
language cache to minimise re-computation, but with a novel cache
admission policy (RUCs).  The language cache is append-only, hence no
synchronisation is required for read-access.  The key difference from
\cite{ValizadehM:seabasreioag}, our use of RUCs, is discussed in \S
\ref{section_data_representation}.

The enumerator has three core parameters.
\begin{itemize}

\item $T$ = maximal number of traces in the specification $(P, N)$.

\item $L$ = number of bits usable for representing each trace from $(P, N)$ in
  memory.

\item $W$ = number of bits $(P, N)$ hashed \EMPH{to} during enumeration.

\end{itemize}
We write \BASE{T}{L}{W} to emphasise those parameters.  Our current
implementation hard-codes all parameters as
\BASE{64}{64}{128}\footnote{For pragmatic reasons, our implementation uses only 126 bits of W=128, and 63 bits
of L = 64, details omitted for brevity.}, but the abstract algorithm
does not depend on this.  The choice of $W = 128$ is a consequence of
the current limitations of WarpCore
\cite{JuengerD:warcoralffhtog,JuengerD:warpcoreImplementation}, a CUDA
library for high-performance hashing of 32 and 64 bit integers.  All
three parameters heavily affect memory consumption. We chose $T = 64$
and $L = 64$ for convenient comparison with existing work in \S
\ref{section_benchmarks}. While $T, L$ and $M$ are parameters of the
abstract algorithm, the implementation is not parameterised: changing
these parameters requires changing parts of the code.  Making the
implementation fully parametric is conceptually straightforward, but introduces a
substantial number of new edge cases, primarily where parameters are
not powers of 2, which increases verification effort.

We now sketch the high-level structure of \CODE{enum}, the entry
point of the enumerator, taking a specification and a cost-homomorphism as arguments.
For ease of presentation, we use LTL formulae
as search space.  Their representation in the implementation is
discussed in \S\ref{section_data_representation}.
\begin{lstlisting}[language=Python]
language_cache = []

def enum(p, n, cost):
    if (p, n) can be solved with Atom then return Atom (*@\label{atom_check}@*)
    language_cache.append([Atom])   (*@\label{atom_init}@*)
    for c in range(cost(Atom)+1, cost(overfit(p, n))):
        language_cache.append([]) 
        for op in [F, U, G, X, And, Or, Not]: (*@\label{all_ltl_ops}@*)
            handleOp(op, p, n, c, cost)
    return overfit(p, n)
\end{lstlisting} Line
\ref{atom_check} checks if the learning problem can be solved with an
atomic proposition. If not, Line \ref{atom_init} initialises the
global language cache with the representation of atomic propositions,
and search starts from the lowest cost upwards.  For each cost $c$ a
new empty entry is added to the language cache. Line \ref{all_ltl_ops}
then maps over LTL connectives and calls \CODE{handleOp} to construct
all formulae of cost $c$ using all suitable lower cost entries in the
language cache.  When no sound formula can be found with cost less
than $\COST{\OVERFIT{P, N}}$, the algorithm terminates, returning
$\OVERFIT{P, N}$. This makes our algorithm \EMPH{complete}, in the
sense of learning a formula for every specification.

\begin{lstlisting}[language=Python,firstnumber=11]
def handleOp(op, p, n, c, cost):
    match op:
        case F:
            for all phi in language_cache(c-cost(F)):    # parallel
                phi_new = branchfree_F(phi)
                relaxedCheckAndCache(p, n, c, phi_new)
        case U:
            for all (cL, cR) in split(c-cost(U)):        # parallel, split(x) gives all (i,j) s.t. i+j = x
                for all phi_L in language_cache(cL):     # parallel
                    for all phi_R in language_cache(cR): # parallel
                        phi_new = branchfree_U(phi_L, phi_R)
                        relaxedCheckAndCache(p, n, c, phi_new)
        case G: ...
        case ...
\end{lstlisting}

\NI The function \CODE{handleOp} dispatches on LTL connectives,
retrieves all previously constructed formulae of suitable cost from
the language cache in parallel (we use \CODE{for all} to indicate
parallel execution), calls the appropriate semantic function, detailed
in the next section, \EG \CODE{branchfree_F} for $\F$, to construct
\CODE{phi\_new}, and then sends it to \CODE{relaxedCheckAndCache} to check if it already
solves the learning task, and, if not, for potential caching.
Most parallelism in our implementation, and the upside of the
language cache's rapid growth, is the concomitant growth in available
parallelism, which effortlessly saturates every conceivable processor.

\begin{lstlisting}[language=Python,firstnumber=25]
def relaxedCheckAndCache(p, n, c, phi_new): # c is a concrete cost
    if phi_new |= (p, n):                   # check if candidate is sound for (p, n) (*@\label{preciseCheck}@*) 
        exit(phi_new)                       # Terminate learning, return phi_new as learned formula
    if relaxedUniquenessCheck(phi_new, language_cache): (*@\label{weakUniquenessCheck}@*)
        language_cache[c].append(phi_new)   # parallel (*@\label{WritePar}@*)

\end{lstlisting}

\NI This last step checks if \CODE{phi\_new} satisfies $(P, N)$. If
yes, the program terminates with the formula corresponding to 
\CODE{phi\_new}.  Otherwise, Line \ref{weakUniquenessCheck} conducts a RUC, a
\EMPH{relaxed} uniqueness check, described in detail in \S
\ref{section_ruc}, to decide whether to cache \CODE{phi\_new}~{}~or not.
Updating the language cache in Line \ref{WritePar} is done in
parallel, and needs little synchronisation, see
\cite{ValizadehM:seabasreioag} for details.  The satisfaction check in
Line \ref{preciseCheck} guarantees that our algorithm is \EMPH{sound}.
It also makes it trivial to implement noisy LTL learning: just replace
the precise check \CODE{phi_new |= (p, n)} with a check that
\CODE{phi_new} gets a suitable fraction of the specification right.

\section{In-memory representation of search space}\label{section_data_representation}

\begin{wrapfigure}{r}{0.35\textwidth}
  \centering
  \includegraphics[width=0.35\textwidth]{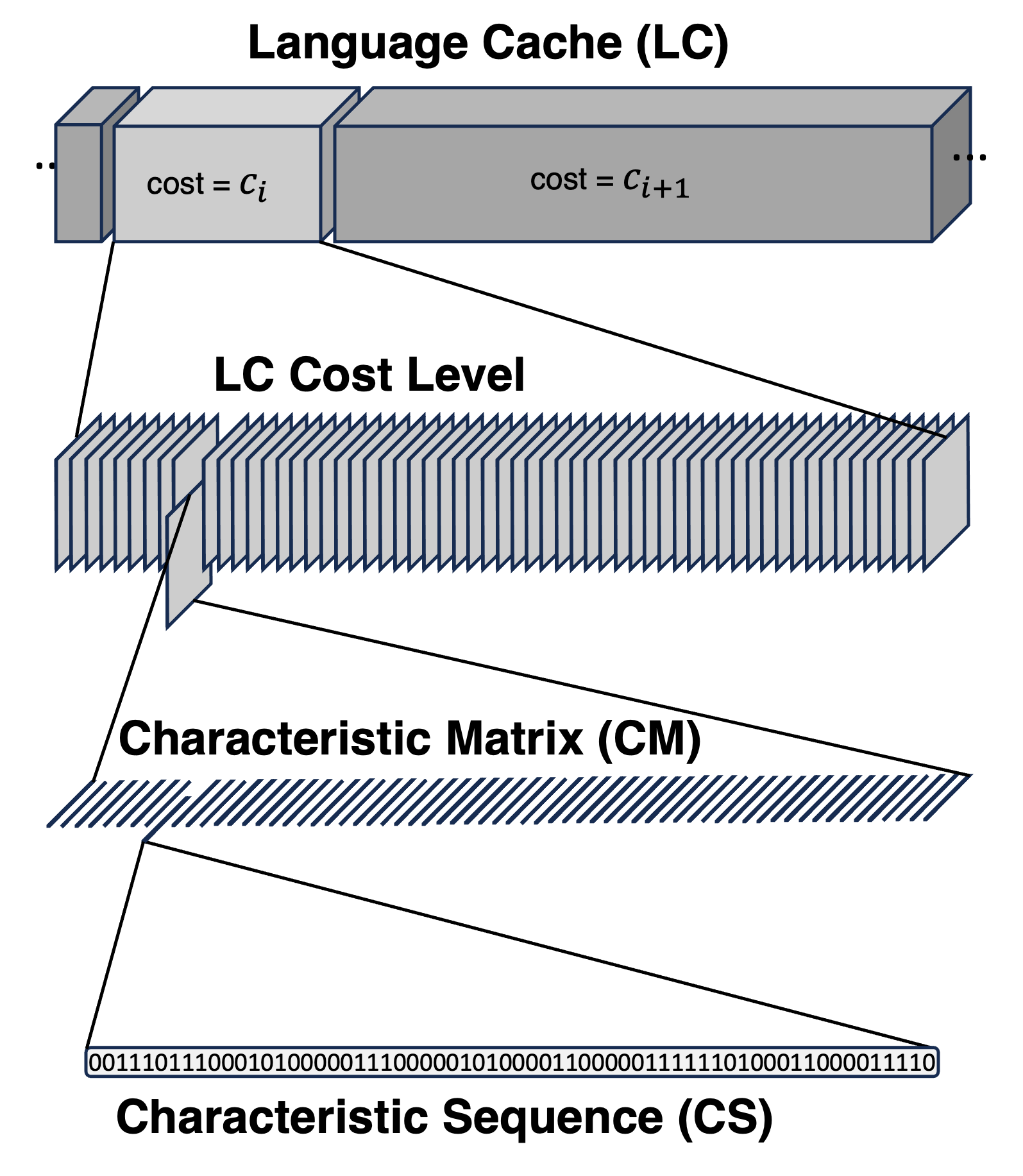}
  \caption{Data representation in memory (simplified).}
  \label{image_allmem}
\end{wrapfigure}

Our enumerator does generate-and-check synthesis. That means
we have two problems: (i) minimising the cardinality of the search
space, \IE the representation of LTL formulae during synthesis; (ii)
making generation and checking as cheap as possible.  For each
candidate $\phi$ checking means evaluating the predicate
\begin{align}\label{ltli_predicate}
  P\subseteq \LANG{\phi} \text{ and } N \cap \LANG{\phi} = \emptyset.
  \tag{$\dagger$}
\end{align}
LTL formulae, the natural choice of search space, suffer
from the redundancies of syntax: every language that is denoted by a
formula at all, is denoted by infinitely many, \EG
$\LANG{\F\phi} = \LANG{\F\F \phi}$. Even 
observational equivalence distinguishes too many formulae: the predicate
(\ref{ltli_predicate}) checks the language of $\phi$ only for elements
of $P \cup N$.  Formulae modulo $P \cup N$ contain exactly the right
amount of information for (\ref{ltli_predicate}), hence minimise the
search space.  However, the semantics of formulae on $P \cup N$ is
given compositionally in terms of the non-empty suffix-closure of $P \cup N$, which
would have to be recomputed at run-time for each new candidate. Since
$P \cup N$ remains fixed, so does $\SUFFIXNE{P \cup N}$, and we can avoid
such re-computation by using formulae modulo $\SUFFIXNE{P \cup N}$ as
search space.  Inspired by \cite{ValizadehM:seabasreioag}, we represent formulae $\phi$
by characteristic functions
$\CHARACTERISTIC{}{\LANG{\phi}} : \SUFFIXNE{P \cup N} \FS \BOOL$, which are implemented as contiguous
bitvectors in memory, but with a twist.  Fix a total order on $P \cup
N$.

\begin{itemize}

\item A \EMPH{characteristic sequence (CS) for $\phi$ over $tr$} is a
  bitvector $cs$ such that $tr, j \models \phi$ iff $cs(j) = 1$.
  For \BASE{64}{64}{128}, CSs are unsigned 64 bit integers.
  
\item A \EMPH{characteristic matrix (CM) representing $\phi$ over $P
  \cup N$}, is a sequence $cm$ of CSs, contiguous in memory,
  such that, if $tr$ is the
  $i$th trace in the order, then $cm(i)$ is the CS for $\phi$ over
  $tr$.

\end{itemize}
See Appendix \ref{appendix_examples} for examples.
This representation has two interesting properties, not present in
\cite{ValizadehM:seabasreioag}: (i) each CS is
\EMPH{suffix-contiguous}: the trace corresponding to $cs(j+1)$ is
the immediate suffix of that at $cs(j)$; (ii) CMs contain
\EMPH{redundancies} whenever two traces in $P \cup N$ share suffixes.
Redundancy is the price we pay for suffix-contiguity.
Figure \ref{image_allmem} visualises our representation in memory.

\PARAGRAPH{Logical operations as bitwise operations}
Suffix-contiguity enables the efficient representation of logical
operations: if $cs = 10011$ represents $\phi$ over the word
$abcaa$, \EG $\phi$ is the atomic proposition $a$, then $\X \phi$ is
$00110$, \IE $cs$ shifted one to the left. Likewise $\neg \phi$
is represented by $01100$, \IE bitwise negation. As we use
unsigned 64 bit integers to represent CSs, $\X$ and negation are executed
as  \EMPH{single} machine instructions! In Python-like pseudo-code:

\begin{minipage}{0.40\linewidth}
\begin{lstlisting}[language=Python]
def branchfree_X(cm):
    return [cs << 1 for cs in cm]
\end{lstlisting}
\end{minipage}
\qquad
\begin{minipage}{0.40\linewidth}
\begin{lstlisting}[language=Python,numbers=right]
def branchfree_Not(cm):
    return [~cs for cs in cm]
\end{lstlisting}
\end{minipage}

\NI Conjunction and disjunction are equally efficient.  More
interesting is $\F$ which becomes the disjunction of shifts by
\EMPH{powers of two}, \IE the number of shifts is logarithmic in the
length of the trace (a naive implementation of $\F$ is linear).  We
call this \EMPH{exponential propagation}, and believe it to be novel
in LTL synthesis\footnote{By representing $\phi$ as a CS, \IE unsigned
integer, we can also read $\F \phi$ as rounding up $\phi$ to the next
bigger power of 2 and then subtracting 1, \CF
\cite{AndersonSE:bittwih}.}:

\begin{minipage}[t]{0.40\linewidth}
\begin{lstlisting}[language=Python]
def branchfree_F(cm):
    outCm = []
    for cs in cm:
        cs |= cs << 1
        cs |= cs << 2
        cs |= cs << 4
        cs |= cs << 8
        cs |= cs << 16
        cs |= cs << 32
        outCm.append(cs)
    return outCm

\end{lstlisting}
\end{minipage}
\qquad
\begin{minipage}[t]{0.40\linewidth}
\begin{lstlisting}[language=Python,numbers=right]
def branchfree_U(cm1, cm2):
    outCm = []
    for i in range(len(cm1)):
        cs1 = cm1[i]
        cs2 = cm2[i]          (*@\label{u_line_0}@*)
        cs2 |= cs1 & (cs2 << 1)   (*@\label{u_line_1}@*)
        cs1 &= cs1 << 1           (*@\label{u_line_2}@*)
        cs2 |= cs1 & (cs2 << 2)   (*@\label{u_line_3}@*)
        cs1 &= cs1 << 2
        cs2 |= cs1 & (cs2 << 4)
        cs1 &= cs1 << 4
        cs2 |= cs1 & (cs2 << 8)
        cs1 &= cs1 << 8
        cs2 |= cs1 & (cs2 << 16)
        cs1 &= cs1 << 16
        cs2 |= cs1 & (cs2 << 32)
        outCm.append(cs2)
    return outCm
\end{lstlisting}
\end{minipage}

\NI See Appendix \ref{appendix_examples} for examples.
To see why this works, note that
$\F \phi$ can be seen as the infinite disjunction $\phi \OR \X\phi \OR
\X^2 \phi \OR \X^3 \phi \OR ...$, where $\X^{n} \phi$ is given by $X^0
\phi = \phi$ and $X^{n+1} \phi = \X \X^n\phi$.  Since we work with
finite traces, $tr, i \not\models \phi$ whenever $i \geq
\SIZE{tr}$. Hence checking $tr, 0 \models \F \phi$ for $tr$ of length
$n$ amounts to checking
\[
tr, 0 \models \phi \OR \X\phi  \OR \X^2 \phi \OR ...  \OR \X^{n-1}\phi
\]
The key insight is that the imperative update \CODE{cs |= cs << j}
propagates the bit stored at $cs(i+j)$ into $cs(i)$ without removing
it from $cs(i+j)$. Consider the flow of information stored in
$cs(n-1)$.  At the start, this information is only at index
$n-1$. This amounts to checking $tr, n-1 \models \X^{n-1}\phi$. Thus
assigning \CODE{cs\ |= cs << 1} puts that information at indices $n-2,
n-1$. This amounts to checking $tr, 0 \models \X^{n-2}\phi \OR
\X^{n-1}\phi$.  Likewise, then assigning \CODE{cs\ |= cs << 2} puts
that information at indices $n-4, n-3, n-2, n-1$.  This amounts to
checking $tr, 0 \models \X^{n-4}\phi \OR \X^{n-3}\phi \OR \X^{n-2}\phi
\OR \X^{n-1}\phi$, and so on. In a logarithmic number of steps, we
reach $tr, 0 \models \phi \OR \X\phi \OR ...  \OR \X^{n-1}\phi$. This
works uniformly for all positions, not just $n-1$. In the limit, this
saves an exponential amount of work over naive shifting.

We can implement $\U$ using similar ideas, with the number of bitshifts also logarithmic in trace length. 
As with $\F$, this works because we can see $\phi \U \psi$ as an infinite disjunction
\[
   \psi \OR (\phi \wedge \X \psi) \OR (\phi \wedge \X (\phi \wedge \X \psi)) \OR \dots
\]
We define (informally) $\phi \U_{\le p} \psi$ as: $\phi$ holds until $\psi$ does within the next $p$ positions,
and $\G_{\ge p} \phi$ if $\phi$ holds for the next $p$ positions.
The additional insight allowing us to implement exponential propagation for $\U$ is to compute both, 
$\G_{\ge 2^i} \phi$ and $\phi \U_{\le 2^i} \psi$, for increasing values of $i$ at the same time.
Appendix \ref{section_correctness_of_bitsmearing} proves correctness of exponential propagation.

In addition to saving work, exponential propagation maps directly to
machine instructions, and is essentially branch-free code for all LTL
connectives\footnote{Including, \EMPH{mutatis mutandis}, past-looking
temporal connectives.}, thus maximises GPU-friendliness of our
learner.  In contrast, previous learners like Flie \cite{NeiderG18},
Scarlet \cite{RahaR:scaanyaflfoltl} and Syslite
\cite{ArifF:syssyngsopffrt}, implement the temporal connectives
naively, \EG checking $tr, i \models \phi \U \psi$ by iterating from
$i$ as long as $\phi$ holds, stopping as soon as $\psi$
holds. Likewise, Flie encodes the LTL semantics directly as a
propositional formula.  For $\U$ this is quadratic in the length of
$tr$ for Flie and Syslite.

\section{Relaxed uniqueness checks}\label{section_ruc}

Our choice of search space, formulae modulo $\SUFFIXNE{P \cup N}$,
while more efficient than bare formulae, still does not prevent the
explosive growth of candidates: uniqueness of CMs is \EMPH{not}
preserved under LTL connectives.  \cite{ValizadehM:seabasreioag}
recommends storing newly synthesised formulae in a ``language cache'',
but only if they pass a uniqueness check.  Without this cache
admission policy, the explosive growth of redundant CMs rapidly swamps
the language cache with useless repetition.  While uniqueness improves
scalability, it just delays the inevitable: there are simply too many
unique CMs. Worse: with \BASE{64}{64}{128}, CMs use up-to 32 times
more memory than language cache entries in
\cite{ValizadehM:seabasreioag}.  We improve memory consumption of our
algorithm by relaxing strictness of uniqueness checks: we allow false
positives (meaning that CMs are falsely classified as being already in
the language cache), but not false negatives.  We call this new cache
admission policy \EMPH{relaxed uniqueness checks} (RUCs).  False
positives mean that less gets cached. False positives are sound: every
formula learned in the presence of false positives is
separating, but no longer necessarily minimal---every minimal solution
might have some of its subformulae missing from the language cache,
hence cannot be constructed by the enumeration.  False positives also
do not affect completeness: in the worst case, our algorithm terminates by
overfitting.

We implement the RUC using non-cryptographic hashing in several steps.
\begin{itemize}

\item We treat each CM as a big bitvector, \IE ignore its internal
  structure.  Now there are two possibilities.
\begin{itemize}

\item The CM uses more than 126 bits. Then we hash it to 126 bits using 
  a variant of MuellerHash from WarpCore.

\item Otherwise we leave the CM unchanged (except padding it with 0s
  to 126 bits where necessary).
  
\end{itemize}
  
\item Only if this 126 bit sequence is unique, it is added to the language cache.
  
\end{itemize}
If the CM is $\leq 126$ bits, then the RUC is precise and 
enumeration performs a full bottom-up enumeration of CMs, so
any learned formula is minimal cost.  This becomes useful in
benchmarking.
\begin{center}
\includegraphics[width=0.75\linewidth]{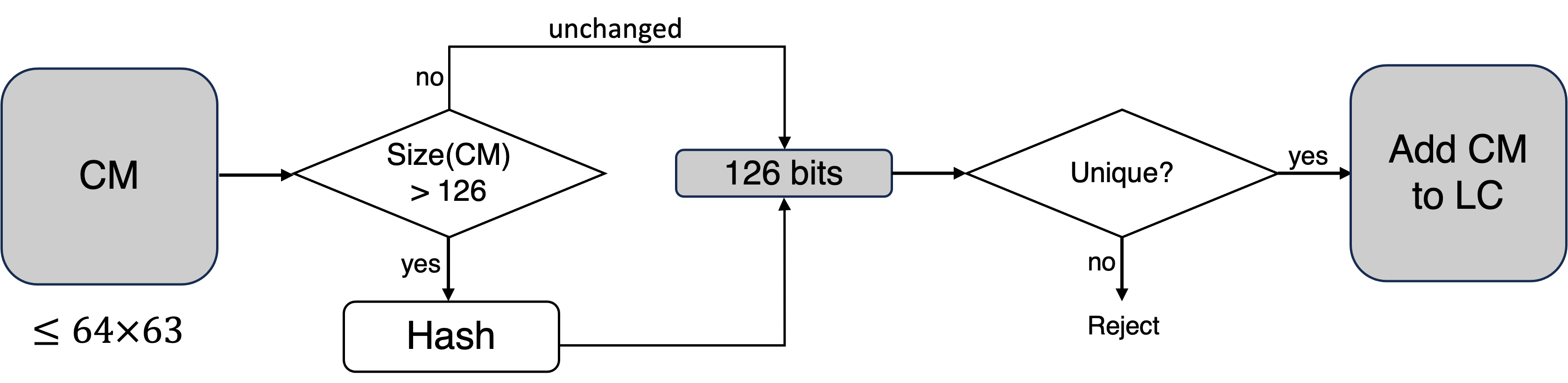}
\end{center}
Note that RUCs implemented by hashing amount to a (pseudo-)random
cache admission policy. See Appendix \ref{appendix_cap} for background
on cache admission polices.  Using RUCs essentially means that hash-collisions
(pseudo-)randomly prevent formulae from being subformulae of any
learned solution. It is remarkable that this is work well in practice, but it
probably means that LTL has sufficient redundancy in formulae
vis-a-vis the probability of hash collisions. We leave a detailed
theoretical analysis as future work.

\section{Divide \& conquer}\label{d_and_c}

The D\&C-unit's job is, recursively, to split specifications until
they are small enough to be solved by \BASE{T}{L}{W} in one go,
and, afterwards recombine the results.  
A naive D\&C-strategy could split $(P, N)$, when needed, into four
smaller specifications $(P_i, N_j)$ for $i, j = 1, 2$, such that $P$
is the disjoint union of $P_1$ and $P_2$, and $N$ of $N_1$ and
$N_2$. Then it learned the $\phi_{ij}$ recursively from
the $(P_i, N_j)$, and finally combine all into
\begin{align*}
  (\phi_{11} \wedge \phi_{12}) \vee (\phi_{21} \wedge \phi_{22})
\end{align*}
which is sound for $(P, N)$, but  is not necessarily minimal. \STARTEG
whenever $\phi_{11}$ implies $\phi_{12}$, then $\phi_{11} \vee (\phi_{21}
\wedge \phi_{22})$ is lower cost\footnote{Such
redundancies can be eliminated, for example, by using theorem
provers.}.  Thus it might be tempting to minimise D\&C-steps.
Alas, the enumerator may run out-of-memory (OOM): the parameters in
\BASE{T}{L}{W} are \EMPH{static} constraints, pertaining to data
structure layout, and do not guarantee successful termination.
Let us call the maximal cardinality $\CARD{(P, N)}$ that
the D\&C-unit sends directly to the enumerator, the \EMPH{split
  window}.  In order to navigate the trade-offs between avoiding OOM
and minimising the approximation ratio, our refined
 D\&C-units below use search to find
as large as possible a split window.  We write $win$ for the
split window parameter. Both implementations split specifications
until they fit into the split window, \IE $\CARD{(P, N)} \leq win$, and
then invoke the enumerator. The split window is then successively
halved, until the enumerator no longer runs OOM but returns a
sound formula.

\PARAGRAPH{Deterministic splitting}
The idea behind $\DETERMNINISTICSPLIT{P, N, win}$ is: if
$\CARD{(P, N)}$ $\leq$ $win$, we send $(P, N)$ directly to the
enumerator. Otherwise, assume $P$ is $\{p_1, ..., p_n\}$. Then $P_1 = \{p_1,
..., p_{n/2}\}$ and $P_2 = \{p_{n/2 + 1}, ..., p_n\}$ are the new
positive sets, and likewise for $N$. (If the specification is given as
two lists of traces, this is deterministic.)  We then make 4 recursive
calls, but remove redundancies in the calls' arguments.
\begin{itemize}
    
  \item  $\phi_{11} = \DETERMNINISTICSPLIT{P_1, N_1, win}$,
  \item  $\phi_{12} = \DETERMNINISTICSPLIT{P_1, N_2 \cap  \LANG{\phi_{11}}, win}$,
  \item  $\phi_{21} = \DETERMNINISTICSPLIT{P_2 \setminus L, N_1, win}$, 
  \item  $\phi_{22} = \DETERMNINISTICSPLIT{P_2 \setminus L, N_2 \cap \LANG{\phi_{21}}, win}$,
    
\end{itemize}
Here $L = \LANG{\phi_{11}} \cup \LANG{\phi_{12}}$.  Assuming that none
of the 4 recursive calls returns OOM, the resulting formula is
$(\phi_{11} \AND \phi_{12}) \OR (\phi_{21} \AND \phi_{22})$.
Otherwise we recurse with $\DETERMNINISTICSPLIT{P, N, win/2}$.

\PARAGRAPH{Random splitting} This variant of the algorithm, written $\RANDOMSPLIT{P, N, win}$, 
is based on the intuition that often a small number
of traces already contain enough information to learn a formula
for the whole specification.  (\STARTEG the 
traces are generated by running the same system multiple times.)
\begin{itemize}
  
  \item $\phi_{11} = \RANDOMAUX{P, N, win}$
  \item $\phi_{12} = \RANDOMSPLIT{P \cap \LANG{\phi_{11}}, N \cap \LANG{\phi_{11}}, win}$
  \item $\phi_{21} = \RANDOMSPLIT{P \setminus \LANG{\phi_{11}}, N \setminus \LANG{\phi_{11}}, win}$
  \item $\phi_{22} = \RANDOMSPLIT{P \setminus \LANG{\phi_{11}}, N \cap \LANG{\phi_{11}}, win}$

\end{itemize}
 The function $\RANDOMAUX{P, N, win}$ first construct a
 sub-specification $(P_{0}, N_{0})$ of $(P, N)$ as follows. Select two
 random subsets $P_{0} \subseteq P$ and $N_{0} \subseteq N$, such that
 the cardinality of $(P_{0}, N_{0})$ is as large as possible but not
 exceeding $win$; in addition we require the cardinalities of $P_{0}$
 and $N_{0}$ to be as equal as possible. Then $(P_0, N_0)$ is sent the
 enumerator. If that returns OOM, $\RANDOMAUX{P, N, win/2}$ is invoked,
 then $\RANDOMAUX{P, N, win/4}, ...$ until the enumerator successfully
 learns a formula.  Once $\phi_{11}$ is available, the remaining
 $\phi_{ij}$ can be learned in parallel.  Finally, we return
 $(\phi_{11} \AND \phi_{12}) \OR (\phi_{21} \AND \phi_{22})$.

 Our benchmarks in \S
\ref{section_benchmarks} show that deterministic and random splitting
display markedly different behaviour on some benchmarks.

\section{Evaluation of algorithm performance}\label{section_benchmarks}

This section quantifies the performance of our implementation. We are
interested in a comparison with existing LTL learners, but also in
assessing the impact on LTL learning performance of different
algorithmic choices.  We benchmark along the following quantitative
dimensions: number of traces the implementation can handle, speed of
learning, and cost of inferred formulae.  In our evaluation we are
facing several challenges.

\begin{itemize}

\item We are comparing a CUDA program running on a GPU with programs,
  sometimes written in Python, running on CPUs.

\item We run our benchmarks in \EMPH{Google Colab Pro}.  It is unclear
  to what extent Google Colab Pro is virtualised.  We observed 
  variations in CPU and GPU running times, for all implementations
  measured.

\item Existing benchmarks are too easy.  They neither force our
  implementation to learn costly formulae, nor terminate later than
  the \EMPH{measurement threshold} of around 0.2 seconds, a minimal
  time the Colab-GPU would take on any task, including toy programs
  that do nothing at all on the GPU.

\item Lack of ground-truth: how can we evaluate the price we pay for
  scale, \IE the loss of formula minimality guarantees from
  algorithmic choices, when we do not know what this minimum is?

\end{itemize}

\PARAGRAPH{Hardware and software used for benchmarking}
Benchmarks below run on \textsc{Google Colab Pro}.  We use Colab Pro
because it is a widely used industry standard for running and
comparing ML workloads.  \textbf{Colab CPU parameters:} Intel Xeon CPU
(``cpu family 6, model 79''), running at 2.20GHz, with 51 GB RAM. \textbf{
  Colab GPU parameters:} Nvidia Tesla V100-SXM2, with System
Management Interface 525.105.17, with 16 GB memory.  We use Python
version 3.10.12, and CUDA version: 12.2.140.  All our timing
measurements are end-to-end (from invocation of $\SYNTH{P, N}$ to its
termination), using Python's \CODE{time} library.

\PARAGRAPH{Benchmark construction}\label{section_benchmark_construction}
Since existing benchmarks for LTL learning are too easy for our
implementation, we develop new ones.  A good benchmark should be
tunable by a small number of explainable parameters that allows users
to achieve hardness levels, from trivial to beyond the
edge-of-infeasibility, and any point in-between. We now describe how
we construct our new benchmarks.

\begin{itemize}

\item 
By $\SIMPLE{\Sigma, k, lo, hi}$ we denote the specifications generated using the
following process: uniformly sample $2\cdot k$ traces from $\{tr \in
\TRACES{\Sigma}\ |\ lo \leq \LENGTH{tr} \leq hi\}$.  Split them into
two sets $(P, N)$, each containing $k$ traces.

\item 
  By $\SCARLET{\Sigma, \phi, k, lo, hi}$ we mean using the sampler
  coming with Scarlet \cite{RahaR:scaanyaflfoltl} to sample
  specifications $(P, N)$ that are separated by $\phi$, where $\phi$ is
  a formula over the alphabet $\Sigma$.  Both, $P$ and $N$, contain
  $k$ traces each, and for each $ tr \in P \cup N$ we have $ lo \leq
  \LENGTH{tr} \leq hi$. The probability distribution Scarlet 
  implements is detailed in Appendix \ref{appendix_scarlet_random}.

\item 
  By $\SAMPLEBENCH{i, k, c}$, where $i, k \in \NAT$ and
  $c \in \{conservative, \neg conservative\}$, we mean the following process, which we
  also call \EMPH{extension by sampling}.
    \begin{enumerate}

    \item Generate $(P, N)$ with $\SIMPLE{\BOOL, i, 2, 5}$.  

    \item\label{def_sample_bench_2} Use our implementation to learn a minimal
      formula $\phi$ for $(P, N)$ that is $\U$-free and in NNF
      (for easier comparison with Scarlet, which can
      neither handle general negation nor $\U$).

    \item Next we sample a specification $(P', N')$ from 
      $\SCARLET{\BOOL, \phi, k, 63, 63}$. 

    \item The final specification is given as follows:
      \begin{itemize}
        
      \item $(P \cup P', N \cup N')$ if $c = conservative$. Hence the final
        specification is a conservative extension of $(P, N)$ and its cost
        is $\COST{P, N}$.
        
      \item $(P', N')$ otherwise.

      \end{itemize}
      
    \end{enumerate}   
    Note that the minimal formula required in Step
    \ref{def_sample_bench_2} exists because for $i \leq 8$, any $(P,
    N)$ generated by $\SIMPLE{\BOOL, i, 2, 5}$ has the property that
    $\CARD{\SUFFIXNE{P \cup N}} \leq 80 < 126$, so our algorithm uses
    neither RUCs nor D\&C, but, by construction, does an exhaustive
    bottom-up enumeration that is guaranteed to learn a minimal sound
    formula.

\item By $\HAMMINGBENCH{\Sigma, l, \delta}$, with $l, \delta \in
  \NAT$, we mean specifications $ (\{tr\}, \HAMMING{tr}{\delta}) $,
  where $tr$ is sampled uniformly from all traces of length $l$ over
  $\Sigma$.

\end{itemize}

\NI Benchmarks from $\SAMPLEBENCH{k, i, c}$ are useful for comparison
with existing LTL learners, and to hone in on specific properties of
our algorithm. But they don't fully address a core problem of using random
traces: they tend to be too easy.  One dimension of ``too easy'' is
that specifications $(P, N)$ of random traces often have tiny
sound formulae, especially for large alphabets.  Hence
we use binary alphabets, the hardest case in this context.  That alone is not enough to force large
formulae.  $\HAMMINGBENCH{\Sigma, l, \delta}$ works well in our
benchmarking: it generates benchmarks that are hard even for the
GPU. We leave a more detailed investigation why as future work.
Finally, in order to better understand the effectiveness of
MuellerHash in our RUC, we use the following deliberately simple map
from CMs to 126 bits.
\begin{quote}

  \textsc{First-k-percent (FKP).} This scheme simply takes the first
  $k\%$ of each CS in the CM. All remaining bits are discarded. The
  percentage $k$ is chosen such that the result is as close as
  possible to 126 bits. \EG for a 64*63 bit CM, $k = 3$.
    
\end{quote}

\PARAGRAPH{Comparison with Scarlet}\label{section_comparison_scarlet}
In this section we compare the performance of our implementation against Scarlet
\cite{RahaR:scaanyaflfoltl}, in order better to understand how much
performance we gain in comparison with a state-of-the-art
LTL learner.  Our comparison with Scarlet is 
implicitly also a comparison with Flie \cite{NeiderG18} and Syslite
\cite{ArifF:syssyngsopffrt} because \cite{RahaR:scaanyaflfoltl}
already benchmarks Scarlet against them, and finds that Scarlet
performs better.  We use the following  benchmarks in our comparison.
\begin{itemize}

\item All benchmarks from \cite{RahaR:scaanyaflfoltl}, which includes older benchmarks for Flie and Syslite.

\item Two new benchmarks for evaluating scalability to high-cost
  formulae, and to high-cardinality specifications.

\end{itemize}
In all cases, we learn $\U$-free formulae in NNF for
easier comparison with Scarlet.  This restriction hobbles our implementation
which can synthesise cheaper formulae in unrestricted LTL.

\begin{TABLE}
  \caption{Comparison of Scarlet with our implementation on existing
    benchmarks. Timeout is 2000 seconds.  On the existing benchmarks
    our implementation never runs OOM or out-of-time (OOT), while Scarlet runs
    OOM in 5.9\% of benchmarks and OOT in 3.8\%.  In computing the
    average speedup we are conservative: we use 2000 seconds whenever Scarlet runs OOT, if
    Scarlet runs OOM, we use the time to OOM.  The ``Lower Cost''
    column gives the percentages of instances where our implementation learns a
    formula with lower cost than Scarlet, and likewise for ``Equal''
    and ``Higher''. Here and below, ``Ave'' is short for the \EMPH{arithmetic mean}. The column on the right reports the average speedup
  over Scarlet of our implementation.}\label{table_existing_work_report}
          {{{\small
\fbox{\begin{tabular}{ccc@{\hspace{1em}}|@{\hspace{1em}}r@{\hspace{1em}}r@{\hspace{1em}}r@{\hspace{1em}}l}

	D\&C & / Hsh	& / Win & Lower Cost & Equal Cost & Higher Cost & Ave Speed-up \\
	
	\cmidrule[1pt]{1-7}
	
	\multirow{6}{*}{\rotatebox[origin=c]{90}{Rand Split}} & \multirow{3}{*}{\rotatebox[origin=c]{90}{FKP}} 
	 	 & 64 & 11.2\%\hspace{0.5cm}	 & 81.9\%\hspace{0.5cm}	 & 6.9\%\hspace{0.5cm}	 &  \textbf{\hspace{0.5cm}\textgreater515x} \\
	 & & 32 & 10.7\%\hspace{0.5cm}	 & 79.8\%\hspace{0.5cm}	 & 9.5\%\hspace{0.5cm}	 & \textbf{\hspace{0.5cm}\textgreater483x} \\
	 & & 16  & 10.9\%\hspace{0.5cm}	 & 76.3\%\hspace{0.5cm}	 & 12.8\%\hspace{0.5cm}	 & \textbf{\hspace{0.5cm}\textgreater320x} \\
	
	\cmidrule{2-7}
	
	& \multirow{3}{*}{\rotatebox[origin=c]{90}{Mueller}} 
	 	 & 64 & 12.3\%\hspace{0.5cm} & 77.9\%\hspace{0.5cm} & 9.8\%\hspace{0.5cm} & \textbf{\hspace{0.5cm}\textgreater58x} \\
	&  & 32 & 11.4\%\hspace{0.5cm} & 72.7\%\hspace{0.5cm} & 15.9\%\hspace{0.5cm} & \textbf{\hspace{0.5cm}\textgreater433x} \\
	&  & 16 & 11.2\%\hspace{0.5cm} & 67.2\%\hspace{0.5cm} & 21.6\%\hspace{0.5cm} & \textbf{\hspace{0.5cm}\textgreater236x} \\
    
	\cmidrule[1pt]{1-7}
    
	\multirow{6}{*}{\rotatebox[origin=c]{90}{Det Split}} & \multirow{3}{*}{\rotatebox[origin=c]{90}{FKP}} 
	 	 & 64& 11.4\%\hspace{0.5cm}	 & 74.4\%\hspace{0.5cm}	 & 14.2\%\hspace{0.5cm}	 & \textbf{\hspace{0.5cm}\textgreater173x} \\
	&  & 32 & 10.4\%\hspace{0.5cm}	 & 70.5\%\hspace{0.5cm}	 & 19.2\%\hspace{0.5cm}	 & \textbf{\hspace{0.5cm}\textgreater103x} \\
	&  & 16 & 10.5\%\hspace{0.5cm}	 & 66.1\%\hspace{0.5cm}	 & 23.3\%\hspace{0.5cm}	 & \textbf{\hspace{0.5cm}\textgreater52x} \\
	
	\cmidrule{2-7}
    
	& \multirow{3}{*}{\rotatebox[origin=c]{90}{Mueller}} 
	 	 & 64 & 12.4\%\hspace{0.5cm}	 & 78.1\%\hspace{0.5cm}	 & 9.5\%\hspace{0.5cm}	 & \textbf{\hspace{0.5cm}\textgreater263x} \\
	&  & 32 & 10.5\%\hspace{0.5cm}	 & 72.5\%\hspace{0.5cm}	 & 16.9\%\hspace{0.5cm}	 & \textbf{\hspace{0.5cm}\textgreater114x} \\
	&  & 16 & 10.5\%\hspace{0.5cm}	 & 66.5\%\hspace{0.5cm}	 & 23.0\%\hspace{0.5cm}	 & \textbf{\hspace{0.5cm}\textgreater46x} \\
	
\end{tabular}}
}
 }}
\end{TABLE}
\PARAGRAPH{Scarlet on existing benchmarks}
We run our implementation in 12 different modes: D\&C by deterministic,
\RESP random splitting, with two different hash functions (MuellerHash
and FKP), and three different split windows (16, 32, and 64).  The
results are visualised in Table \ref{table_existing_work_report}.  We
make the following observations.  On existing benchmarks, our implementation usually
returns formulae that are roughly of the same
cost as Scarlet. They are typically only larger on benchmarks with a
sizeable specification, \EG 100000 traces, which forces our implementation
into D\&C, with the concomitant increase in approximation ratio due to the cost of recombination. However the
traces are generated by sampling from trivial formulae (mostly $\F p,
\G p$ or $\G \neg p$).  Scarlet handles those well.
\cite{RahaR:scaanyaflfoltl} defines a parameterised family
$\phi^n_{seq}$ that can be made arbitrarily big by letting $n$ go to
infinity. However in \cite{RahaR:scaanyaflfoltl} $n < 6$, and even
on those Scarlet run OOM/OOT, while our implementation handles all in a short
amount of time.  On 
existing benchmarks, our implementation runs on average at least 46 times faster.  We
believe that this surprisingly low worst-case average speedup is largely because the
existing benchmarks are too easy, and the timing measurements are
dominated by GPU startup latency.  The comparison on harder benchmarks
below shows this.

\PARAGRAPH{Scarlet and high-cost specifications}\label{section_high_cost_comparison_with_scarlet}
The existing benchmarks can all be solved with small formulae.
This makes it difficult to evaluate how our implementation scales
when forced to learn high-cost formulae. In order to ameliorate this
problem, we create a new benchmark using $\HAMMINGBENCH{\BOOL, l,
  \delta}$ for $l = 3, 6, 9, ..., 48$ and $\delta = 1, 2$.  We
benchmark with the aforementioned 12 modes.  The left of Table
\ref{table_high_cost_comparison_with_scarlet} summarises the results.
\begin{TABLE}
  \caption{On the left, comparison between Scarlet and our
    implementation on $\HAMMINGBENCH{\BOOL, l, \delta}$ benchmarks
    with $l = 3, 6, 9, ..., 48$ and $\delta = 1, 2$. Timeout is 2000
    sec. Reported percentage is fraction of specifications that were
    successfully learned.  On the right, comparison on benchmarks from
    $\SAMPLEBENCH{5, 2^k, conservative}$ for $k = 3, 4, 5, ...,
    17$. All benchmarks were run to conclusion, Scarlet's OOMs occurred between
    1980.21 sec for $(2^{17}, 2^{17})$, and 16568.7 sec for $(2^{13},
    2^{13})$.}
  \label{table_high_cost_comparison_with_scarlet}
  \label{figure_high_card_comparison_with_scarlet}
  \begin{minipage}[t]{0.49\linewidth}
    {\small
\fbox{\begin{tabular}{ccc|rr}

	D\&C & / Hsh	& / Win & Delta=1 & Delta=2 \\
	
	\cmidrule[1pt]{1-5}
	
	\multirow{6}{*}{\rotatebox[origin=c]{90}{Rand Split}} & \multirow{3}{*}{\rotatebox[origin=c]{90}{FKP}} 
		& 64 & 100\%\hspace{0.2cm}	 & 100\%\hspace{0.2cm}	\\
	& & 32 & 100\%\hspace{0.2cm}	& 100\%\hspace{0.2cm}	\\
	& & 16  & 100\%\hspace{0.2cm}	 & 100\%\hspace{0.2cm}	\\
	
	\cmidrule{2-5}
	
	& \multirow{3}{*}{\rotatebox[origin=c]{90}{Mueller}} 
		 & 64 & 100\%\hspace{0.2cm}	 & 75\%\hspace{0.2cm}	\\
	&  & 32 & 100\%\hspace{0.2cm}	& 75\%\hspace{0.2cm}	\\
	&  & 16 & 100\%\hspace{0.2cm}	 & 100\%\hspace{0.2cm}	\\
	
	\cmidrule[1pt]{1-5}
	
	\multirow{6}{*}{\rotatebox[origin=c]{90}{Det Split}} & \multirow{3}{*}{\rotatebox[origin=c]{90}{FKP}} 
		 & 64 & 100\%\hspace{0.2cm}	 & 100\%\hspace{0.2cm}	\\
	&  & 32 & 100\%\hspace{0.2cm}  & 100\%\hspace{0.2cm}	\\
	&  & 16 & 100\%\hspace{0.2cm}	& 100\%\hspace{0.2cm}	\\
	
	\cmidrule{2-5}
	
	& \multirow{3}{*}{\rotatebox[origin=c]{90}{Mueller}} 
		 & 64 & 100\%\hspace{0.2cm}	 & 88\%\hspace{0.2cm}	\\
	&  & 32 & 100\%\hspace{0.2cm}	& 100\%\hspace{0.2cm}	\\
	&  & 16 & 100\%\hspace{0.2cm}	& 100\%\hspace{0.2cm}	\\
	
	\cmidrule[1pt]{1-5}
	
	\multicolumn{3}{c|}{Scarlet} & \textbf{7\%}\hspace{0.2cm} & \textbf{7\%}\hspace{0.2cm} \\
	
\end{tabular}}
}

  \end{minipage}
  \begin{minipage}[t]{0.49\linewidth}
    {\small
\fbox{\begin{tabular}{c|cccc}
		
		(\CARD{P}, \CARD{N}) & \hspace{2mm} & \makecell{Our impl. \\ Time (Cost)} & \hspace{2mm} & \makecell{Scarlet\\Time (Cost)}  \\
		
		\cmidrule[1pt]{1-5}
		
		$(2^{3}, 2^{3})$ && 0.31s (12) && 1532.85s (19) \\
		$(2^{4}, 2^{4})$ && 0.32s (12) && 1463.67s (17) \\
		$(2^{5}, 2^{5})$ && 0.36s (12) && 2867.47s (17) \\
		$(2^{6}, 2^{6})$ && 0.34s (12) && 5691.98s (17) \\
		$(2^{7}, 2^{7})$ && 0.63s (20) && OOM \\
		$(2^{8}, 2^{8})$ && 0.95s (19) && OOM \\
		$(2^{9}, 2^{9})$ && 0.72s (19) && OOM \\
		$(2^{10}, 2^{10})$ && 1.09s (19) && OOM \\
		$(2^{11}, 2^{11})$ && 1.32s (19) && OOM \\
		$(2^{12}, 2^{12})$ && 1.66s (19) && OOM \\
		$(2^{13}, 2^{13})$ && 2.46s (19) && OOM \\
		$(2^{14}, 2^{14})$ && 4.62s (20) && OOM \\
		$(2^{15}, 2^{15})$ && 8.35s (19) && OOM \\
		$(2^{16}, 2^{16})$ && 15.52s (19) && OOM \\
		$(2^{17}, 2^{17})$ && 30.49s (19) && OOM \\
		
\end{tabular}}
}
  
  \end{minipage}
\end{TABLE}
This benchmark clearly shows that Scarlet is mostly unable
to learn bigger formulae, while our implementation handles all swiftly.

\PARAGRAPH{Scarlet and high-cardinality specifications}\label{section_high_card_comparison_with_scarlet}
The previous benchmark addresses scalability to high-cost
specifications. The present comparison with Scarlet seeks to quantify
an orthogonal dimension of scalability: high-cardinality
specifications.  Our benchmark is generated by $\SAMPLEBENCH{i, 2^k,
  conservative}$ for $i = 5$, and $k = 3, 4, 5, ..., 17$.
Using $i = 5$ ensures getting a few concrete times from Scarlet rather
than just OOM/OOT; $k \leq 17$ was chosen as Scarlet's sampler makes
benchmark generation too time-consuming otherwise.
The choice of parameters also ensures that the cost of each
benchmark is moderate $(\le 20)$. This means any difficulty with
learning arises from the sheer number of traces. Unlike the previous
two benchmarks, we run our implementation only in one configuration: using
MuellerHash, and random splitting with window size 64 (the difference
between the variants is too small to affect the comparison with
Scarlet in a substantial way). The results are also presented on the
right of Table \ref{figure_high_card_comparison_with_scarlet}.  This
benchmark clearly shows that we can handle specifications at least
2048 times larger, despite Scarlet having approx.~3 times more memory
available. Moreover, not only is our implementation much faster and can
handle more traces, it also finds substantially smaller formulae in
all cases where a comparison is possible.

\PARAGRAPH{Hamming benchmarks}\label{benchmark_hamming}
We have already used $\HAMMINGBENCH{...}$ in our comparison with
Scarlet. Now we abandon existing learners, and delve deeper into the
performance of our implementation by having it learn costly formulae. This
benchmark is generated using $\HAMMINGBENCH{\BOOL, l, \delta}$ for
$l = 3, 6, 9, ..., 48$ and $\delta = 1, 2$. As above, the implementation learns 
$\U$-free formulae in NNF.  Table \ref{label_hamming_figures}
gives a more detailed breakdown of the results. The uniform cost of
overfitting on each benchmark is given for comparison:
\begin{FIGURE}
  \includegraphics[width=1\linewidth]{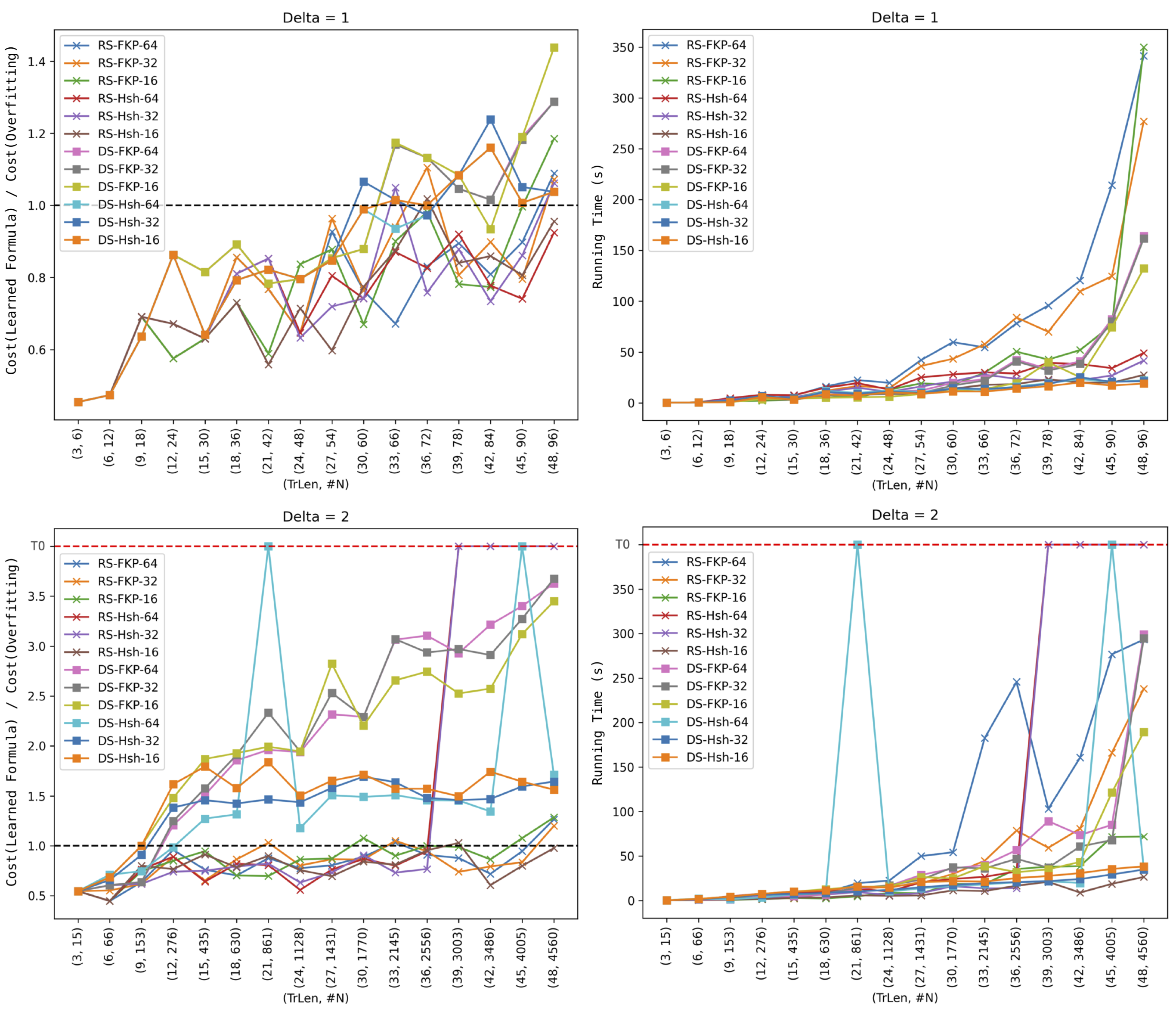}        
  \caption{Here RS means random-splitting, DS
    deterministic-splitting. The numbers 16, 32, 64 are the used
    splitting window. Hsh is short for MuellerHash.  The x-axis is
    annotated by $(trLen, \#N)$, giving the length of the single trace in $P$,
    and the cardinality $\# N$ of $N$. TO
     denotes timeout. Timeout is 2000 seconds.  On the left, the y-axis gives
    the ratio $\frac{\text{cost of learned formula}}{\text{cost of
        overfitting}}$, the dotted line at 1.0 is the cost of overfitting.}
  \label{label_hamming_figures}
\end{FIGURE}
{\small
\begin{center}
\begin{tabular}{ |c|c|c|c|c|c|c|c|c|c|c|c|c|c|c|c|c| }
  \hline
  Length of $tr$  &   3 & 6 & 9 & 12 & 15 & 18 & 21 & 24 & 27 & 30 & 33 & 36 & 39 & 42 &  45 & 48  \\  \hline
  Cost of  overfitting &  22 &  38 &  55 &  73 &  92 &  111 &  129 &  147 &  164 &  182 &  201 &  219 &  238 &  256 &  274 &  292  \\ \hline
\end{tabular}
\end{center}
} \NI This benchmark shows the following. Hamming benchmarks are hard for our
implementation, and sometimes run for $> 3$ minutes: we successfully
force our implementation to synthesise large formulae, and that has an effect
on running time.  The figures on the left show that random splitting
typically leads to smaller formulae in comparison with deterministic
splitting, especially for $\delta = 2$. Indeed, we may be seeing a
sub-linear increase in formula cost (relative to the cost of
overfitting) for random splitting, while for deterministic splitting,
the increase seems to be linear. In contrast, the running time of the
implementation seems to be relatively independent from the splitting
mechanism.  It is also remarkable that the maximal cost we see is only
about 3.5 times the cost of overfitting: the algorithm processes $P$
and $N$, yet over-fitting happens only on $P$, which contains a single
trace.  Hence the cost of over-fitting (292 in the worst case) is not
affected by $N$, which contains up to 4560 elements (of the same length
as the sole positive trace).

\begin{TABLE}
  \caption{All run times are below the measurement
    threshold.}\label{table_RUC_data} {\small
\fbox{\begin{tabular}{c|cc|cc}
		
		 \multicolumn{1}{c}{} & \multicolumn{2}{c}{FKP} &  \multicolumn{2}{c}{MuellerHash} \\
		
		\cmidrule(rl){2-3} \cmidrule(rl){4-5}
		
		(\CARD{P}, \CARD{N}) & \hspace{0.2cm}AveExtraCost\hspace{0.2cm} & \hspace{0.2cm}OOM\hspace{0.2cm} &  \hspace{0.5cm}AveExtraCost\hspace{0.2cm} & \hspace{0.2cm}OOM\hspace{0.2cm} \\
		
		\cmidrule[1pt]{1-5}
		
		(8, 8) 		 & 0.0\%	& 0.0\%		& 0.0\%		& 0.0\% \\
		
		\cmidrule[0.2pt]{1-5}
		
		(12, 12) 	& 2.4\%		& 12.8\%	& 1.9\%		& 24.5\% \\ 
		(16, 16) 	& 4.1\%		& 19.3\%	& 0.7\%		& 32.6\% \\
		(20, 20) 	& 4.1\%		& 22.4\%	& 0.4\%		& 32.9\% \\
		(24, 24) 	& 3.9\%		& 24.0\%	& 0.1\%		& 33.6\% \\
		(28, 28) 	& 3.6\%		& 24.9\%	& 0.0\%		& 32.9\% \\ 
		(32, 32) 	& 2.7\%		& 26.3\%	& 0.2\%		& 40.3\% \\
		
\end{tabular}}
}

\end{TABLE}

\PARAGRAPH{Benchmarking RUCs}\label{benchmarks_RUC}
Our algorithm uses RUCs, a novel cache admission policy, and it is
interesting to gain a more quantitative understanding of the effects
of (pseudo-)randomly rejecting some CMs. We cannot hope to come to a
definitive conclusion here.  Instead we simply compare MuellerHash
with FKP, which neither distributes values uniformly across the hash
space (our 126 bits) to minimise collisions, nor has the avalanche
effect where a small change in the input produces a significantly
different hash output. This weakness  is valuable for
benchmarking because it indicates how much a hash function can degrade
learning performance. (Note that for the edge case of specifications that can be separated from just the first $k\%$ alone, FKP should perform better, since it leaves the crucial bits unchanged.)
The benchmark data is generated with $\SAMPLEBENCH{8, 24,
  conservative}$. Table \ref{table_RUC_data} summarises our
measurements. We note the following.  The loss in formula cost is
roughly constant for each hash: it stabilises to around
$0.2\%$ for MuellerHash, and a little above $2.5\%$ for FKP. Hence
MuellerHash is an order of magnitude better.  Nevertheless, even
$2.5\%$ should be irrelevant in practice, and we conjecture that
replacing MuellerHash with a cryptographic hash will have only a
moderate effect on learning performance.  A surprising number of
instances run OOM, more so as specification size grows, with
MuellerHash more than FKP. We leave a detailed understanding
of these phenomena as future work.

\PARAGRAPH{Masking}\label{section_RUC_masking}
The previous benchmarks suggested that naive hash functions like FKP
sometimes work better than expected. Our last benchmark seeks to
illuminate this in more detail and asks: can we relate the information
loss from hashing and the concomitant increase in formula cost? A
precise answer seems to be difficult. We run a small experiment: after
MuellerHashing CMs of size 64*63 bits to 126 bits, we add an
additional information loss phase: we \EMPH{mask out} $k$ bits, \IE we
set them to 0. This destroys all information in the $k$ masked bits.
After masking, we run the uniqueness check.  We sweep over
 $k = 1, ..., 126$ with stride 5 to mask out benchmarks generated with $\SAMPLEBENCH{8,
  32,  \neg conservative}$.  Figure \ref{label_RUC_masking_figures} shows the
results.  Before running the experiments, the authors expected a
gradual increase of cost as more bits are masked
out. Instead, we see a phase transition when approx.~75 to 60 bits are
not masked out: from minimal cost formulae before, to OOM/OOT
after, with almost no intermediate stages. Only a tiny number of instances have 1 or 2 further cost
levels between these two extremes.  We leave an explanation of this
surprising behaviour as future work.

\begin{FIGURE}
  \includegraphics[width=0.8\linewidth]{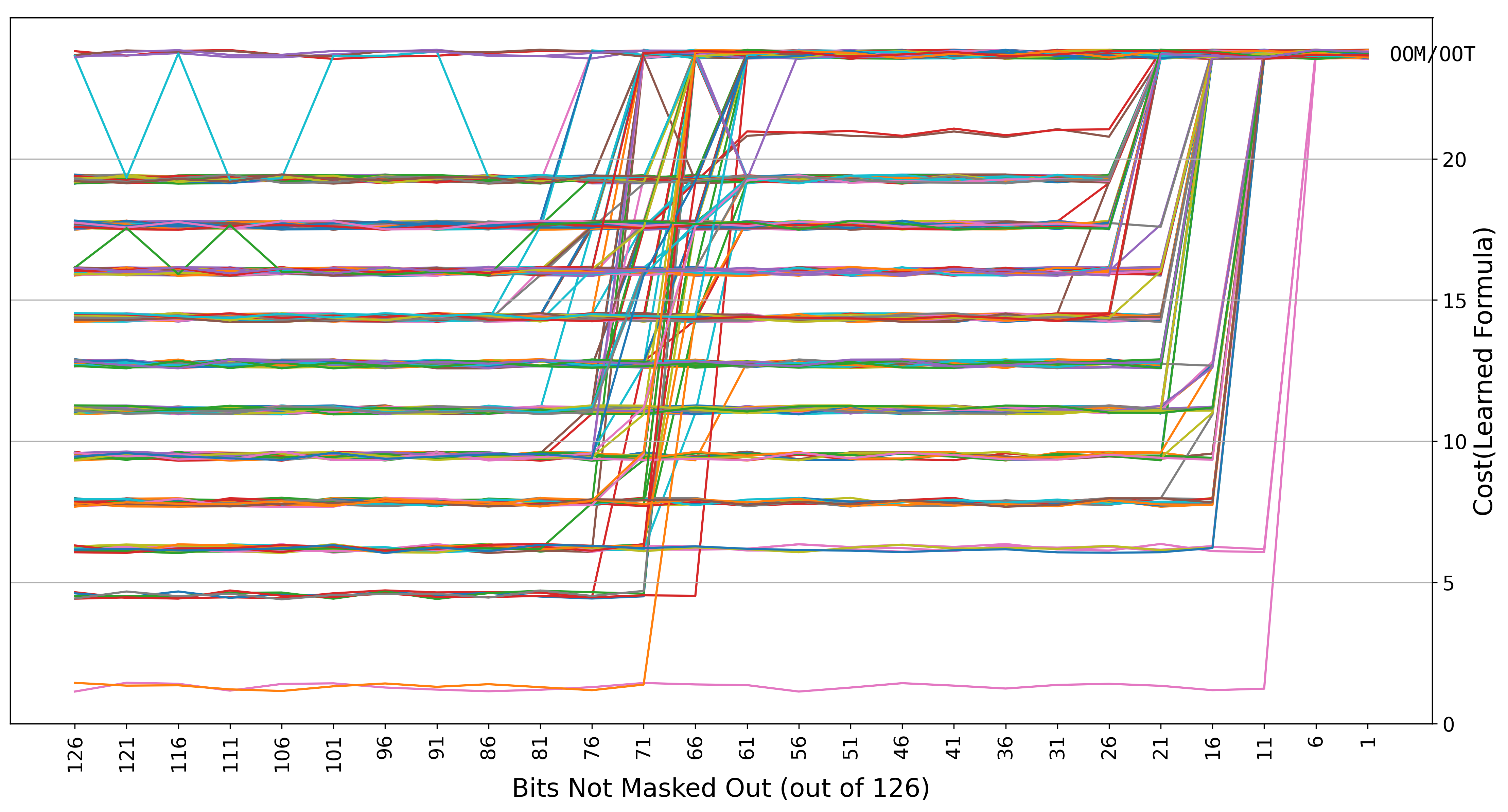}    
  \caption{Effects of masking on formula cost. Timeout is 200
    seconds. Colours correspond to different $(P, N)$. The slight 'wobble' on all graphs is deliberately
    introduced for readability, and is not in the data.}\label{label_RUC_masking_figures}
\end{FIGURE}

\section{Conclusion}

The present work demonstrates the effectiveness of carefully tailored
algorithms and data structures for accelerating LTL learning on GPUs.
We close by summarising the reasons why we achieve scale: high degree
of parallelism inherent in generate-and-test synthesis; application of
divide-and-conquer strategies; relaxed uniqueness checks for (pseudo-)randomly curtailing
the search space; and succinct, suffix-contiguous data representation,
enabling exponential propagation where LTL connectives map directly to
branch-free machine instructions with predictable data movement.  All
but the last are available to other learning tasks that have suitable
operators for recombination of smaller solutions.
\begin{quote}
\EMPH{LTL and GPUs, a match made in heaven}.
\end{quote}

\newpage
\bibliographystyle{splncs04}
\bibliography{bib/semantic, bib/ltl_learning}

\newpage
\appendix
\section{Worked examples of CS manipulation by bitshifts}\label{appendix_examples}

In this section we illustrate constructing and using CSs by examples.

\PARAGRAPH{Characteristic sequences}
Consider the atomic proposition $g$ where $g \in \Sigma$ is a character.
Here are some examples of CS for formulae $\X^n g$ over the word
``squeegee''.
\[
\begin{array}{rcc}
  \phi &\quad \qquad& \text{$cs$} \\ \hline
 g  &&  00000100  \\
 \X g  &&  00001000  \\
 \X \X g  &&  00010000  \\
 \X \X \X g  &&  00100000  \\
 \X \X \X \X g  &&  01000000  \\
 \X \X \X \X \X g  &&  10000000  \\
 \X \X \X \X \X \X g  &&  00000000  \\
 \X \X \X \X \X \X \X g  &&  00000000  \\
\end{array}
\]
The next example is also CSs over ``squeegee'', but now for formulae of the
form $\bigvee_{n} \X^n g$, which are closely related to $\F$.
\[
\begin{array}{rcc}
  \phi &\quad \qquad& \text{$cs$} \\ \hline
 g  &&  00000100  \\
 g \OR \X g  &&  00001100  \\
 g \OR \X g \OR \X^{2}  g  &&  00011100  \\
 g \OR \X g \OR \X^{2}  g \OR \X^{3} g  &&  00111100  \\
 g \OR \X g \OR \X^{2}  g \OR \X^{3} g \OR \X^{4} g  &&  01111100  \\
 g \OR \X g \OR \X^{2}  g \OR \X^{3} g \OR \X^{4} g \OR \X^{5} g  &&  11111100  \\
 g \OR \X g \OR \X^{2}  g \OR \X^{3} g \OR \X^{4} g \OR \X^{5} g \OR \X^{6} g  &&  11111100  \\
 g \OR \X g \OR \X^{2}  g \OR \X^{3} g \OR \X^{4} g \OR \X^{5} g \OR \X^{6} g \OR \X^{7} g  &&  11111100  \\
\end{array}
\]

\PARAGRAPH{Exponential propagation for $\F$ and $\U$}
To see exponential propagation in action consider a formula $\phi$
that is represented by $cs$ over some trace.  We begin by computing
the CS for $\F \phi$ with exponential propagation. In order to
visualise how the CSs are changed by assignments, we use the notation
$\{pre\}\ P\ \{post\}$ from Hoare logic. It should be read as: if,
before executing of program $P$, the state is correctly described by
$pre$, then, after $P$ has executed, the assertion $post$ holds.
Assume the initial state is $cs = 0000000000000100$, which is a CS of
length 16. Then we compute the CS for $\F \phi$, as given by (the body
of) \CODE{branchfree_F}~{}~with the following steps.
\[
{\small
\begin{array}{rcrcllcl}
  \{ cs = 0000000000000100 \} &\quad& \mathtt{cs}  &|=& \mathtt{ cs \ll 1 } &\quad & \{ cs = 0000000000001100 \} \\
  \{ cs = 0000000000001100 \} && \mathtt{cs}  &|=& \mathtt{ cs \ll 2 } && \{ cs = 0000000000111100 \} \\
  \{ cs = 0000000000111100 \} && \mathtt{cs}  &|=& \mathtt{ cs \ll 4 } && \{ cs = 0000001111111100 \} \\
  \{ cs = 0000001111111100 \} && \mathtt{cs}  &|=& \mathtt{ cs \ll 8 } && \{ cs = 1111111111111100 \} \\
\end{array}
}
\]
At the end of this computation we reach $cs = 1111111111111100$ and
that is the CS for $\F \phi$, as required.  In order to see
exponential propagation for $\U$ in action, let us assume we are given
two CSs, $cs_1 = 11111110$ which represents $\phi$ and $cs_2 =
00000001$ representing $\psi$, both over some trace of length 8. We
construct the CS representing $\phi \U \psi$.  We carry out
exponential propagation as given by (the body of) \CODE{branchfree_U}~{}~with
the following steps.
\[
{\small
\begin{array}{rcrcllcl}
   \{ cs_1 = 11111110, cs_2 = 00000001\}  &\quad& \mathtt{cs2} &|=& \mathtt{ cs1 \& (cs2 \ll 1)}  &\quad & \{ cs_1 = 11111110, cs_2 = 00000011 \} \\
   \{ cs_1 = 11111110, cs_2 = 00000011  \}  && \mathtt{cs1 }&\&=& \mathtt{ (cs1 \ll 1)}        && \{ cs_1 = 11111100, cs_2 = 00000011 \} \\
   \{  cs_1 = 11111100, cs_2 = 00000011  \}  && \mathtt{cs2} &|=& \mathtt{ cs1 \& (cs2 \ll 2)}  && \{ cs_1 = 11111100, cs_2 = 00001111 \} \\
   \{  cs_1 = 11111100, cs_2 = 00001111  \}  && \mathtt{cs1 }&\&=& \mathtt{ (cs1 \ll 2)}        && \{ cs_1 = 11110000, cs_2 = 00001111 \} \\
   \{  cs_1 = 11110000, cs_2 = 00001111  \}  && \mathtt{cs2} &|=& \mathtt{ cs1 \& (cs2 \ll 4)}  && \{ cs_1 = 11110000, cs_2 = 11111111 \} \\
\end{array}
}
\]
Now $ cs_2 = 11111111$ is the CS for $\phi \U \psi$, as required.

\section{Correctness and complexity of branch-free semantics for temporal operators}\label{section_correctness_of_bitsmearing}

We reproduce here a slightly more general version (for any length) of the exponentially propagating algorithms for computing $\F$ and $\U$.

\begin{minipage}[t]{0.40\linewidth}
\begin{lstlisting}[language=Python]
def branchfree_F(cs):
    L = len(cs)
    for i in range(log(L)+1): 
        cs |= cs << 2**i
    return cs

\end{lstlisting}
\end{minipage}
\qquad
\begin{minipage}[t]{0.40\linewidth}
\begin{lstlisting}[language=Python,numbers=right]
def branchfree_U(cs1, cs2):
    L = len(cs1)
    for i in range(log(L)+1):
        cs2 |= cs1 & (cs2 << 2**i)  
        cs1 &= cs1 << 2**i
    return cs2
\end{lstlisting}
\end{minipage}

\NI They are lifted to CMs pointwise. As a warm-up, let us start with $\F$.

\begin{lemma}
Let $cs$ be the characteristic sequence for $\phi$ over a trace of length $L$. 
The algorithm above computes the characteristic sequence for $\F \phi$.
Assuming bitwise boolean operations and shifts by powers of two have unit costs,
the complexity of the algorithm is $O(\log(L))$.
\end{lemma}

To ease notations, let us introduce $\F_{\le p}$ with the semantics
$tr, j \models \F_{\le p} \phi$ if there is $j \leq k < \min(j + p, \LENGTH{tr})$ with $tr, k \models \phi$.

\begin{proof}
Let us write $cs$ for the characteristic sequence for $\phi$ over $tr$, and 
$cs_i$ for the characteristic sequence after the $i${th} iteration. We write $\log(x)$ as a shorthand for $\lfloor \log_{2}(x) \rfloor$.
We write $\F cs$ for $\F \phi$, and $\F_{\le p} cs$ for  $\F_{\le p}\ \phi$.
We show by induction that for all $i \in [0, \log(L)+1]$, for all $tr$ of length $L$ we have:
\[
\forall j \in [0, L],\ tr, j \models cs_i \Longleftrightarrow tr, j \models \F_{\le 2^i} cs.
\]
This is clear for $i = 0$, as it boils down to $cs_0 = cs$.
Assuming it holds for $i$, by definition $cs_{i+1}(j) = cs_i(j) \vee (cs_i \ll 2^i)(j) = cs_i(j) \vee cs_i(j + 2^i)$, hence 
\[
\begin{array}{lll}
tr, j \models cs_{i+1} 
& \Longleftrightarrow & 
tr, j \models cs_i, \text{\ or\ }\ tr, j + 2^i \models cs_i \\
& \Longleftrightarrow & 
tr, j \models \F_{\le 2^i} cs, \text{\ or\ }\ tr, j + 2^i \models \F_{\le 2^i} cs \\
& \Longleftrightarrow &
tr, j \models \F_{\le 2^{i+1}} cs.
\end{array}
\]
This concludes the induction proof.
For $i = \log(L)$ we obtain
\[
\forall j \in [0, L],\ tr, j \models cs_i \Longleftrightarrow tr, j \models \F_{\le L} cs \Longleftrightarrow tr, j \models \F cs,
\]
since clearly $\F = \F_{\le L}$, when restricted to traces not exceeding $L$ in length.
\end{proof}

\NI We now move to $\U$. 

\begin{lemma}
Let $cs_1,cs_2$ the characteristic sequences for $\phi_1$ and $\phi_2$, both over traces of length $L$. 
The algorithm above computes the characteristic sequence for $\phi_1 \U \phi_2$.
Assuming bitwise boolean operations and shifts by powers of two have unit costs,
the complexity of the algorithm is $O(\log(L))$.
\end{lemma}

Again to ease notations, let us introduce $\U_{\le p}$ with the semantics
$tr, j \models \phi_1 \U_{\le p} \phi_2$ if there is $j \leq k < \min(j + p, \LENGTH{tr})$ 
such that $tr, k \models \phi_2$ and for all $i \leq k' < k$ we have $tr, k' \models \phi_1$.
We will also need $\G_{\ge p}$ defined with the semantics
$tr, j \models \G_{\ge p} \phi$ if for all $j \leq k < \min(j + p, \LENGTH{tr})$ we have $tr, k \models \phi$.

\begin{proof}
Let us write $cs_{1,i}$ and $cs_{2,i}$ for the respective characteristic sequences after the $i${th} iteration.
We show by induction that for all $i \in [0, \log(L)+1]$, for all $tr$ of length $L$, for all $j \in [0, L]$,
we have:
\begin{itemize}
\item $tr, j \models cs_{1,i} \Longleftrightarrow tr, j \models \G_{\ge 2^i} cs_1$, and
\item $tr, j \models cs_{2,i} \Longleftrightarrow tr, j \models cs_1 \U_{\le 2^i} cs_2.$
\end{itemize}
This is clear for $i = 0$, as it boils down to $cs_{1,0} = cs_1$ and $cs_{2,0} = cs_2$.
Assume it holds for $i$. 
Let us start with $cs_{1,i+1}$: 
by definition
\begin{align*}
  cs_{1,i+1}(j) &= cs_{1,i}(j) \wedge (cs_{1,i} \ll 2^i)(j)\\
                &= cs_{1,i}(j) \wedge cs_{1,i}(j + 2^i),
\end{align*}
hence it is the case that
\[
\begin{array}{lll}
tr, j \models cs_{1,i+1} 
& \Longleftrightarrow & 
tr, j \models cs_{1,i}, \text{\ and \ }\ tr, j + 2^i \models cs_{1,i} \\
& \Longleftrightarrow & 
tr, j \models \G_{\ge 2^i} cs_1, \text{\ and \ }\ tr, j + 2^i \models \G_{\ge 2^i} cs_1 \\
& \Longleftrightarrow &
tr, j \models \G_{\ge 2^{i+1}} cs_1.
\end{array}
\]
Now, by definition $cs_{2,i+1}(j) = cs_{2,i}(j) \vee (cs_{1,i}(j) \wedge (cs_{2,i} \ll 2^i)(j))$, which is equal to
$cs_{2,i}(j) \vee (cs_{1,i}(j) \wedge cs_{2,i}(j + 2^i))$. 
Hence 
\[
\begin{array}{lll}
tr, j \models cs_{2,i+1} 
& \Longleftrightarrow & 
tr, j \models cs_{2,i}, \text{or\ }\ (tr, j \models cs_{1,i}, \text{\ and \ }\ tr, j + 2^i \models cs_{2,i}) \\
& \Longleftrightarrow & 
tr, j \models cs_1 \U_{\le 2^i} cs_2, \text{\ or \ }\ \\
& & \left( tr, j \models \G_{\ge 2^i} cs_1, \text{\ and \ }\ tr, j + 2^i \models cs_1 \U_{\le 2^i} cs_2 \right) \\
& \Longleftrightarrow &
tr, j \models cs_1 \U_{\le 2^{i+1}} cs_2.
\end{array}
\]
This concludes the induction proof.
For $i = \log(L)$ we obtain
\[
\forall j \in [0, L],\ tr, j \models cs_{2,i} \Longleftrightarrow tr, j \models cs_1 \U_{\le L} cs_2 \Longleftrightarrow tr, j \models cs_1 \U cs_2,
\]
since clearly $\U = \U_{\le L}$ for all sufficiently short traces.
\end{proof}

\section{Related work: cache admission policies}\label{appendix_cap}

Caches are one of the most widely used concepts in computer science
for increasing a system's performance. Caches reserve memory to store data
for future retrieval, a space-time trade-off.  The fundamental issue
with caching is thus dealing with the limitations of the available
memory. There are two main points in time where this is done:
\begin{itemize}

\item Before caching: a cache \EMPH{admission} policy (CAP) decides if
  an item is worth caching.
  
\item After caching: a cache \EMPH{replacement} policy (CRP) decides
  which existing entry need to go to make space for a new one.
  
\end{itemize}
They are not mutually exclusive. Both have in common that they seek to
make a lightweight prediction on whether a piece of data will be
likely used again soon in order to decided what to admit / replace.
In hardware processors, the CAP is usually trivial: cache
everything\footnote{An exception are processors with NUMA caches where
the CAP question ``\EMPH{should} this be placed in the cache?'' is
refined to ``\EMPH{where} in the cache should this be placed?''  so as
to minimise the time required to transport cached data to the
processor element that needs it. In addition, modern processors have
dedicated \EMPH{prefetching} components \cite{MittalS:surrecptfpc},
which analyse the stream of memory reads, detect patterns in it (\EG
if memory access to addresses $x$, $x+8$, $x+16$, $x+24$ becomes
apparent, a prediction is made that the next few memory reads will be
to $x+32$, $x+40$, $x+48$ and so on) and request data items at addresses  extending  this
pattern ahead of time, which places them in the cache.}.  This is based on
the empirical observation of \EMPH{temporal locality}: for most
programs, the fact that a piece of data is used now is a reasonable
predictor for it being used again soon.  In contrast with their simple
CAPs, processors use sophisticated CRPs, including
least-recently-used, most-recently-used, FIFO, or LIFO. All require
that the cache associates meta-data with cached elements, and do not
make sense as CAP.  The only exception is the (pseudo-)random CRP.

The situation is quite different in network-based caching, such as
\EMPH{content delivery networks} like Akamai, which cache frequently
requested data to make the internet faster (see, \EG
\cite{KirilinV:rlcleabcafcd}). Here the CAP is usually based on the
fact that network requests on the internet adhere to some reasonably
well understood probability distribution (\EG Zipf-like) and it is
reasonably cheap to decide where in the distribution a candidate for
cache entry sits.

To the best of our knowledge, our RUC of (pseudo-)randomly rejecting
potential CMs, is new.  This is because, unlike the caching discussed
above, each CM is accessed equally often by future stages of the
algorithm, so predicting how often a cache entry will be used is
pointless. The relevant question in LTL learning is: does this CM play a
necessary role in a minimal (or low-cost) LTL formula? Answering this question is
hard, probably as hard as as LTL learning itself, hence infeasible as
a CAP. Understanding better why our (pseudo-)random CAP is effective
is an interesting question for future work, and the surprising
benchmarks in \S \ref{section_RUC_masking} might be taken to indicate
that more radical CAPs might be viable.

\newcommand{\PARESY}{\textsc{Paresy}}

\section{Related work: detailed comparison with \PARESY{}  \cite{ValizadehM:seabasreioag}}\label{appendix_comparison}

\PARESY{} \cite{ValizadehM:seabasreioag}, the first GPU-accelerated
regular expression inferencer, was the main influence on the present
work and it is important to clarify how both relate.  Table
\ref{lable_comparison_pldi_cav} discusses the key points.  In summary,
both have a similar high-level structure (\EG bottom-up enumeration,
language cache, bitvector representation), but are completely
different in low-level details (\EG staging, redundant representation,
shift-based). This opens interesting avenues for further work. It is
clear that regular expression inference, and indeed any form of search
based program synthesis can benefit from D\&C and from RUCs, albeit at
the cost of loosing minimality guarantees.  A more complex question
is: can the costs arising from the redundant representation of
suffixes be avoided, for example by a staged guide table for LTL
connectives, similar to the staged guide table in
\cite{ValizadehM:seabasreioag}, and what would be the effects on
performance of such a dramatic change of algorithm be?

\begin{table}[t]
\centering
\caption{Comparison of \PARESY{} \cite{ValizadehM:seabasreioag} and
  the present work regarding core algorithmic
  features.}\label{lable_comparison_pldi_cav}\label{tab:my_label}
\begin{tabular}{@{}l|l|l@{}}
\toprule
Feature                                             & \PARESY{}                         & Present paper                       \\ \midrule
GPU-based                                           & Yes                               & Yes                                 \\ \hline
What is learned                                     & Regular expressions               & LTL formulae                        \\ \hline
Generate-and-test                                   & Yes                               & Yes                                 \\ \hline
Generation                                          & Bottom-up enumeration             & Bottom-up enumeration               \\ \hline
Exhaustive                                          & Yes                               & No                                  \\ \hline
Source of incomplete generation                     & N/A                               & RUC                                 \\ \hline
Closure                                             & Infix                             & Suffix                              \\ \hline
Search space                                        & $(P, N$) quotiented by            & $(P, N)$ quotiented by              \\
                                                    & infix-closure                     & suffix-closure                      \\ \hline
Irredundant representation                          & Yes                               & No                                  \\
of candidates                                       &                                   &                                     \\ \hline
Bitvector representation                            & Yes                               & Yes                                 \\ \hline
Suffix-contiguous                                   & No                                & Yes                                 \\ 
representation of candidates                        &                                   &                                     \\ \hline
Max size of bitvector                               & 126 bits                          & 64*63 bits                          \\ 
in given implementation                             &                                   &                                     \\ \hline
Language cache                                      & Yes                               & Yes                                 \\ \hline
Cache admission policy                              & Precise uniqueness check          & Relaxed uniqueness check            \\ 
                                                    &                                   & (false positives allowed)           \\ \hline
D\&C                                                & No                                & Yes                                 \\ \hline
Semantics of candidates                             & Staged with guide table           & Exponentially propagated            \\
                                                    &                                   & bitshifts                           \\ \hline
Staging \WRT $(P, N)$                               & High (\EG guide table)            & Low                                 \\ \hline
Scalable                                            & No                                & Yes                                 \\ \hline
Sound                                               & Yes                               & Yes                                 \\ \hline
Complete                                            & Yes                               & Yes                                 \\ \hline
Minimality guarantees                               & Yes                               & No                                  \\ \hline
Causes of non-minimality                            & N/A                               & RUC and D\&C                        \\ 
\bottomrule
\end{tabular}
\end{table}

\section{Scarlet's sample generation}\label{appendix_scarlet_random}

Our $\SAMPLEBENCH{i, k, c}$ is built on Scarlet's sampler
\cite{RahaR:scaanyaflfoltl} which implements the algorithm proposed in
\cite{Bernardi:linalgftrsfrl}. We now sketch how it works. For
sampling positive traces, proceed as follows.

\begin{itemize}

\item Given a suitable formula, convert the formula into an equivalent
  DFA.
  
\item Sampling starts from the initial state of the DFA.

\item From a state, sample uniformly at random a transition, register
  the character, and move to the corresponding target state of the
  transition.

\item If the state is accepting, there's a non-zero probability to
  stop (uniform, counted as a transition). Otherwise proceed with
  sampling more characters.

\end{itemize}
For negative traces, the same algorithm is used, albeit on the
complement DFA.

\end{document}